\documentclass[pra,onecolumn,superscriptaddress,preprintnumbers,amsmath,amssymb,nofootinbib,floatfix,longbibliography]{revtex4-2}
\usepackage{graphicx,bm}
\usepackage{float}
\usepackage{subfigure}
\usepackage{amsmath}
\usepackage{cases}
\usepackage[overload]{empheq}
\usepackage{amsfonts}
\usepackage{amsthm, amssymb,bbm,bm}
\usepackage[utf8]{inputenc}
\usepackage[english]{babel}
\usepackage{xcolor}
\usepackage{bbold}
\usepackage{graphicx}
\usepackage{appendix}

\newcommand{\DN}[1]{\textcolor{black}{#1}}

\begin{document}
\title{Supersolidity of polariton condensates in photonic crystal waveguides }
\author{Davide Nigro}
\affiliation{Dipartimento di Fisica, Universit\`{a} di Pavia, via Bassi 6, I-27100 Pavia, Italy}
%\email{davide.nigro@unipv.it}
 \author{Dimitrios Trypogeorgos}
 \affiliation{CNR Nanotec, Institute of Nanotechnology, via Monteroni, 73100, Lecce, Italy}
 \author{Antonio Gianfrate}
 \affiliation{CNR Nanotec, Institute of Nanotechnology, via Monteroni, 73100, Lecce, Italy}
  \author{Daniele Sanvitto}
 \affiliation{CNR Nanotec, Institute of Nanotechnology, via Monteroni, 73100, Lecce, Italy}
 \author{Iacopo Carusotto}
 \affiliation{INO-CNR Pitaevskii BEC Center and Dipartimento di Fisica, Universit\`a di Trento, 38123 Povo, Italy}
 \author{Dario Gerace}
 \affiliation{Dipartimento di Fisica, Universit\`{a} di Pavia, via Bassi 6, I-27100 Pavia, Italy}

\begin{abstract}
Condensation of exciton-polaritons has been recently observed in gap-confined states of one-dimensional photonic crystal waveguides. Here we focus on the theoretical emergence of a second emission threshold in this platform, in addition to the one associated with condensation at zero-momentum, due to the nonlinear polariton scattering from the condensate into finite momentum eigenmodes. The physics of this spatially modulated condensate is related to a spontaneous breaking of both phase and translational symmetries simultaneously, bearing similarities with the highly sought supersolid phase in Helium and ultracold atomic gases but with a novel mechanism typical of the driven-dissipative scenario. We then propose clear-cut and unequivocal experimental signatures that would allow to identify such non-equilibrium supersolidity of polariton condensates.
\end{abstract}

\maketitle

\section{Introduction}. 
{Exciton-polaritons are elementary excitations arising from the strong coupling between photonic and excitonic eigenmodes in low-dimensional semiconductor nanostructures \cite{Sanvitto2016}. Such low-dimensional polariton excitations inherit the best of two worlds: the low photon effective mass, and the strong exciton nonlinearity, making them appealing candidates for applications in  nonlinear photonics \cite{Sturm2014All-opticalInterferometer,Suarez-Forero2021EnhancementInteractions,Datta2022HighlyMoS2}, and, on a more fundamental level, for studies of quantum fluid behaviors, such as condensation and superfluidity in a novel non-equilibrium scenario that is peculiar of driven-dissipative systems \cite{Kasprzak2006,Amo2009,Amo2011,Carusotto_Ciuti_RMP2013,Fontaine_Nature_KPZ,Bloch_nat_rev_phys2022}. Intense efforts have been later directed towards laterally confining microcavity polaritons, with the aim of 
achieving condensation and superfluidity at lower thresholds \cite{Yoon2022EnhancedPotentials,Sun2017,Schneider2017Exciton-polaritonEngineering,Ferrier2011InteractionsCondensates,Wertz20210np}. 
In this context, photonic crystal waveguides with embedded excitonic materials have been proposed as promising candidates to engineer exciton-polariton dispersion and losses~\cite{GeracePRB2007,HaMy2020,Zanotti2022}.
Low-threshold polariton condensation has been recently achieved in one-dimensional photonic crystal (1DPC) waveguides with embedded quantum wells (QWs) \cite{Ardizzone2022PolaritonContinuum}, following earlier indications of condensation in gap solitons of fully etched microcavity samples with in-plane periodicity~\cite{Baboux2018,Tanese2013}.  
It has been later shown that such polariton condensation is facilitated by the creation of an effective potential well for negative mass polaritons induced by the exciting laser spot \cite{Riminucci2023prl}.
Hence, theoretical models of out-of-equilibrium condensation in a multi-band exciton-photon coupled system with symmetry-dependent losses have emphasized the crucial role played by the simultaneous presence of negative mass and energy gap in reaching the condensation threshold \cite{Nigro2023prb,SigurdssonNguyenNguyen2024}.}

\begin{figure}[t]
    \centering
    \includegraphics[width=\textwidth]{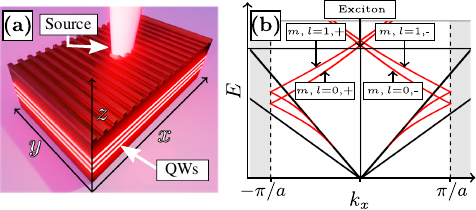}
    \caption{$\textbf{(a)}$: Sketch of the proposed setup. A planar dielectric waveguide is vertically pumped by a non-resonant continuous wave (cw) source. The structure supports QWs along $z$, and it is assumed to be uniform in the transverse direction, $y$. The top surface is characterized by a corrugation with period $a$ along $x$. 
    $\textbf{(b)}$: Sketch of the relevant energy-wavevector ($E$ vs $k_x$) dispersions along $x$, folded in the first Brillouin zone ($k_x \in [-\pi/a,\pi/a]$). The thick horizontal line indicates the nearly flat exciton resonance. Black oblique lines are the light lines of the waveguide core and the upper cladding (air). Red solid lines describe a pair of photonic modes ($m=0,1$) supported by the planar waveguide, which can be further separated in propagating ($+$, positive slope) and counter-propagating ($-$, negative slope) branches around $k_x=0$. }
    \label{fig:1}
\end{figure}

 In the present Letter we focus on power regimes above the polariton condensation threshold, exploring the interplay between condensation and nonlinear polariton scattering processes. We theoretically anticipate the existence of a second emission threshold favouring the scattering from the polariton condensate at zero momentum to equal energy side-branches at finite wave vector. Under suitable conditions, the resulting steady state configuration displays peculiar features that can be understood in terms of a spontaneous breaking of phase and translational symmetries occurring simultaneously. The former is attributed to the onset of condensation in an off-resonant driving scenario, while the latter is evidenced by the presence of peculiar spatially modulated patterns. In particular, such a behavior is hereby assessed as an emergent supersolid state, an elusive phase of matter that has sparked widespread interest in the last decade, especially in connection with the long-standing controversy about its observation in Helium~\cite{Boninsegni2012rmp} and its recent realization in ultracold atomic clouds~\cite{Leonard2017nature,TanziPRL2019_prima_osservazione,Bottcher2019,ChomazPRX2019,Recati_nat_rev_phys2023}. Here we provide a novel playground to explore the physics of such an exotic condensed matter state in the context of driven-dissipative systems, establishing the polariton supersolidity as the supersolid state emerging from a non-equilibrium polariton condensate. 
 % \DN{Given the subtlety of the concept of superfluidity in non-equilibrium systems \cite{PhysRevLett.105.020602}, in particular for condensates in gap states \cite{Grudinina2023}, ongoing work is being devoted to assess the superfluid properties of this new state of matter. In this respect, our theoretical investigation (see e.g. the SM file, "Instability analysis") reveals that modulations spontaneously created on top of the polariton condensate are unconditionally subsonic, thus not harming the superfluid nature of the system.} 
% % , also taking into account the persistence of metastable supercurrents that only depend on the long-range superfluid coherence and not on the Bogoliubov dispersion \cite{PhysRevLett.105.020602}.\\

\section{Theoretical background}
\begin{figure*}[ht]
    \centering
    \includegraphics[width=\linewidth]{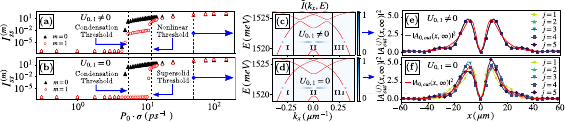}
    \caption{(\textbf{a},\textbf{b}) Steady state photonic emission, $I^{(m)}_{ss}$ (for $m=0,1$, respectively), as a function of the injection rate, $P_0\cdot \sigma$ ($\sigma=10\,\mu m$).(\textbf{c},\textbf{d}) The spectrally resolved emission normalized to the peak intensity in momentum space, i.e. $\tilde{I}(k_x,E)=I(k_x,E)/max(I(k_x,E))$, is shown for $P_0\cdot \sigma=50$ ps$^{-1}$. Emission is normalized to each peak in zones (I,III) and (II), respectively. The actual ratio between peak emission in regions (I,III) and (II) is about 6. Solid red lines in each panel display the calculated polariton bands. (\textbf{e},\textbf{f}) Spatial behavior of 5 independent realizations ($j=1,\ldots,5$) of the total steady state emission pattern in real space, $\vert A^{(j)}_{out}(x,\infty)\vert^2$,
     %=\lim_{t\to \infty} \vert A^{(j)}_{out}(x,\infty)\vert^2 $ 
     compared to that associated to the condensate component only ($m=0$), i.e. $\vert A_{0,out}(x,\infty)\vert^2$, for $P_0\cdot \sigma=50$ ps$^{-1}$ ($\sigma=5 \mu m$). 
     %for comparison to Ref.~\cite{Riminucci2023prl}). 
     The values of all the model parameters used in these simulations are provided in the End Matter section.}
    \label{fig:2}
\end{figure*}

The setup under investigation is sketched  in Fig.~\ref{fig:1}a.
%and described in the caption. 
It has been already shown that polariton condensation, and the peculiar features observed in steady state emission patterns, can be captured by an effective driven-dissipative 1D theory in this case~\cite{Riminucci2023prl,Nigro2023prb}, as a consequence of single photonic and  excitonic dispersions along the periodicity direction ($x$). Here, the physics close to normal incidence, i.e., $k_x=0$, at polariton condensation and beyond is explored by means of a generalized theoretical framework, considering the interplay between the non-resonant source, i.e., $P=P_0\exp{\{-(x^2)/(2\sigma^2)\}}$, two reservoir densities, four photonic linear modes $A_{m,l}\equiv A_{m,l}(x,t)$, and the corresponding four replicas of an exciton resonance $X_{m,l}\equiv X_{m,l}(x,t)$. As shown in Fig.~\ref{fig:1}b, these four photonic modes are understood as approximating the actual dispersion of two guided modes of the structure ($m=0$ and $m=1$), which can be further separated in propagating ($l=+$, positive slope) and counter-propagating ($l=-$, negative slope) branches around $k_x=0$. We notice that the use of four excitonic replicas is based on the observation that each $A_{m,l}$ mode is characterized by its own field profile and polarization, which is independently coupled to the excitonic field. Hence, at mode matching, the photonic branch $(m,l)$ is effectively coupled to the corresponding replica, $X_{m,l}$. In a short hand notation, we will henceforth refer to these photonic/excitonic branches as components of a 8-dimensional vector, $\vec{\psi}=\vec{\psi}(x,t)$. {Examples of the polariton dispersion resulting from the strong coupling of photonic and excitonic modes are shown in Figs.~\ref{fig:2}(c,d)}. Further details on the 1D theory employed in this work are reported in the End Matter section, including all the parameter values used in our numerical simulations. Further technical information is reported in the SM section. Here, we first focus on the property that is most relevant to understand the supersolid phase transition in this context, and the results reported in the following section.

In fact, the theoretical analysis reported in the present Letter is focused on two main coupling contributions of the mixed exciton-photon fields: a nonlinear term, which we formally express as $g_{xx}\vec{W}[\vec{\psi}]$, and a linear one, i.e.,  $H_{0,1}\vec{\psi}$. In the former, $g_{xx}$ denotes the exciton-exciton nonlinearity, i.e., it accounts for scattering processes between pairs of exciton modes. This is evident by noticing that $\vec{W}[\vec{\psi}]$ can be regarded as the functional derivative of the following contact potential~\cite{Greiner1996}
{\small
\begin{equation}\label{eq:nonlinear_coupling}
\mathcal{U}=\int dx \left(X^*_{0,+}X^*_{0,-}X_{1,+}X_{1,-} + c.c. \right) \, ,
 \end{equation}
}
which captures Hamiltonian contributions  associated to the conversion of an exciton pair  $(X_{1,+},X_{1,-})$ into the pair $(X_{0,+},X_{0,-})$, and vice versa. 
%In particular, such term is responsible for the existence of a second threshold above condensation. 
The term $H_{0,1}\vec{\psi}$ captures, instead, the possible coherent coupling introduced by the 1DPC potential between $(m,\pm)$ and $(m^{'},\mp)$ photonic branches. In particular, in this case one can interpret this term as a functional derivative, i.e., $H_{0,1}\vec{\psi}=\delta \mathcal{H}_{0,1}/\delta \vec{\psi}^*$, in which
 {\small 
 \begin{equation}\label{eq:linear_coupling}
      \mathcal{H}_{0,1}=\hbar U_{0,1}\int dx(A_{0,+}^{*}A_{1,-}+A_{0,-}^{*}A_{1,+} + c.c),
\end{equation}}
such that the energy scale $U_{0,1}$ quantifies the strength of such a linear coupling. \\
%\DN{As shown in the following, the presence of a non-zero $U_{0,1}$ not only affects the behavior at condensation threshold, but also and most importantly the one at the nonlinear threshold triggered by $g_{xx}\vec{W}[\vec{\psi}]$}.
In the following, we will show that $U_{0,1}\neq 0$ not only affects the behavior at condensation threshold, but also (and most importantly) the one at the nonlinear scattering threshold into $(m=1,l=\pm)$ modes that is triggered by $g_{xx}\vec{W}[\vec{\psi}]$. Before showing any result from numerics, let us now anticipate an interpretation that is solely based on symmetry arguments, which are at the basis of our definition of the phase transition to a non-equilibrium  supersolid. \\
First, we notice that, since the equations of motion are invariant under a global phase $U(1)$ symmetry, the following phase symmetry holds
{\small
 \begin{equation}\label{eq:phase_condensate}
     \vec{\psi}\to \vec{\psi}\,^{'}=e^{i\theta} \vec{\psi}, \, \theta \in [-\pi,\pi]  \, .
 \end{equation}
}
However, when $U_{0,1}=0$ the equations of motion are characterized by an extra $U(1)$ symmetry, i.e., 
{\small
\begin{subequations}
\label{eq:phase_symmetry_m1} 
\begin{eqnarray}
A_{1,+}&\to A^{'}_{1,+}=e^{i \phi_{L} }A_{1,+},\,\, X_{1,+}&\to X^{'}_{1,+}=e^{i \phi_{L} }X_{1,+} \\
A_{1,-}&\to A^{'}_{1,-}=e^{i \phi_{R} }A_{1,-},\,\, X_{1,-}&\to X^{'}_{1,-}=e^{i \phi_{R} }X_{1,-}
\end{eqnarray}
\end{subequations}
}
for any possible $\phi_{L}\in [-\pi,\pi]$, provided $\phi_{L}+\phi_{R}=0$.\\
As it is known, under non-resonant pumping the order parameter $\vec{\psi}$ becomes non-zero at polariton condensation threshold, and the former $U(1)$ symmetry is spontaneously broken~\cite{Carusotto_Ciuti_RMP2013}. When $U_{0,1}=0$, also the symmetry described in Eq.~\ref{eq:phase_symmetry_m1} spontaneously breaks upon further increase of the input power, thus resulting in a steady state configuration that breaks both phase and translational symmetry, which is the polariton supersolid.\\

\section{Results} The onset of polariton condensation is typically identified by characterizing the intensity and the spectrum of the light emitted by the sample, the latter being accessible in energy-momentum resolved measurements. In our framework, such observables correspond to the steady-state integrated emission associated to the $m$-th photonic component, i.e., $I^{(m)}_{ss}= \lim_{t\to \infty}\int\,\mbox{d}x\, \vert A_{m,out}(x,t)\vert^2 \,$, and the Fourier transform of the output field, $A_{out}(x,t)$, performed over a finite time window deep in the steady-state regime, i.e., $I(k_x,E)=\vert A_{out}(k_x,E)\vert^2$ (see, e.g., Ref.~\cite{Nigro2023prb}). By assuming a linear coupling to the environment (as in input-output theory), we define the total output field, $A_{out}(x,t)$, as the spatial interference of the output fields associated to each photonic component, i.e., $A_{out}(x,t)=\sum_{m}A_{m,out}(x,t)$, with $A_{m,out}(x,t) \propto\sum_{l=\pm} A_{m,l}(x,t)$.

We first pay attention to the behavior of integrated emission as a function of the injection rate, i.e., $P_{0}\cdot \sigma$. Results for $U_{0,1}\neq 0 $ and $U_{0,1}=0$ are reported in Figs.~\ref{fig:2}a and \ref{fig:2}b, respectively. Features displayed by data are compatible with the scenario described in the previous section. On increasing the injection rate, independently of the value of $U_{0,1}$, a clear threshold marking the onset of polariton condensation is visible around $P_{0}\cdot \sigma \simeq 5-6$ps$^{-1}$ (black triangles, $m=0$ data). However, when $U_{0,1}\neq 0$, the tiny emission signal $I^{(1)}_{ss}$ present in panel Fig.~\ref{fig:2}a at condensation threshold and above evidences that some population is coherently transferred to $m=1$ modes. If one further increases the injection rate, a second threshold appears slightly above $P_{0}\cdot \sigma \simeq 10$ps$^{-1}$, as evidenced by data for $m=1$. In particular, since such a trend in data is present both in Fig.~\ref{fig:2}a and Fig.~\ref{fig:2}b, we conclude it is a nonlinear effect triggered by exciton-exciton interactions $g_{xx}\vec{W}[\vec{\psi}]$. Such a fact is also supported by numerical results obtained for $U_{0,1}=0$ and $g_{xx}=0$, which show no emission from $m=1$ modes at any injected power, as reported in Fig.~S1 of the SM section.

\begin{figure}[t]
    \centering
    \includegraphics[width=\textwidth]{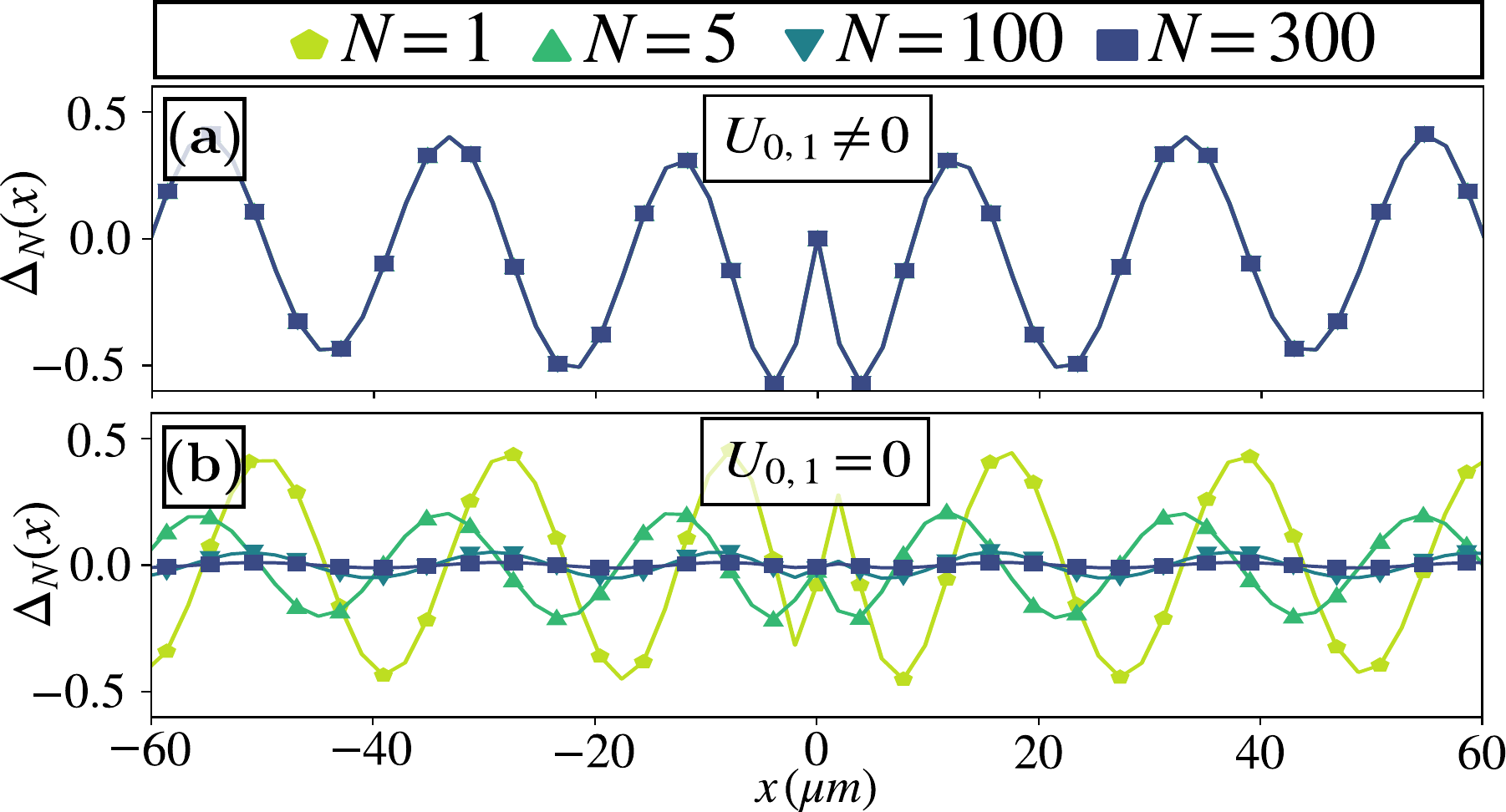}
    \caption{Spatial behavior of the mean modulation pattern, $\Delta_{N}(x)$, calculated for different values of the total number of configurations, $N$ (top legend), and for (\textbf{a}) $U_{0,1}\neq 0$ and (\textbf{b}) $U_{0,1}=0$. Other parameters are set as in Fig.~\ref{fig:2}e,f.}\label{fig:3}
\end{figure}

We now pay attention to the behavior above the second threshold in the two cases. In particular, from now on we set $P_0\cdot \sigma= 50\,ps^{-1}$. We first consider steady-state energy-momentum resolved measurements. Results are shown in Figs.~\ref{fig:2}c and \ref{fig:2}d. In both cases the emission profile is characterised by three distinct regions $k_x$ direction. The flat emission in region (II), which is dark at exactly $k_x=0$, is related to the polariton condensate, and it has been discussed elsewhere \cite{Ardizzone2022PolaritonContinuum,Riminucci2023prl,Nigro2023prb,SigurdssonNguyenNguyen2024}. However, the two peaks in regions (I) and (III) are a distinctive feature of the nonlinear scattering from the condensate into to $m=1$ polaritons states at the same energy, with well defined wavevectors along $x$. In particular, since the peaks in regions (I) and (III) are located at $k_x=\pm\bar{k}_x$, with $\bar{k}_x\simeq 0.3$ $\mu$m$^{-1}$, we may expect they give rise to periodic modulations in real space.\\
In fact, this is confirmed by results in Figs.~\ref{fig:2}e and \ref{fig:2}f, in which we report the outcome of 5 independent realizations ($j=1,\cdots,5$) of the equilibration process in the presence of the random noise $\vec{\xi}\equiv \vec{\xi}(x,t)$ for $U_{0,1}\neq 0$ and $U_{0,1}=0$, respectively. We notice that a 
stochastic noise is required in order to detect the eventual presence of spontaneously broken symmetries ~\cite{Altland2010}. %, i.e., time evolution towards the steady state at fixed external conditions. 
In these two panels, we compare the total steady state emission profiles, $\vert A_{\,out}(x,\infty)\vert^2= \lim_{t\to \infty}\vert A_{\,out}(x,t)\vert^2$, to the condensate emission profile, $\vert A_{0,\,out}(x,t\to \infty)\vert^2$ (red solid line). In particular, for comparison with the spatial emission profiles in Ref.~\cite{Riminucci2023prl}, we used a cw source with $\sigma=5 \mu m$. Independently of the value of $U_{0,1}$, we notice that the total emission extends in space by following the polariton condensate profile ($m=0$). However, for $U_{0,1}\neq 0$ there exists a unique steady state configuration (trajectories for different $j$ are perfectly superimposed). On the contrary, the independent spatial patterns obtained for $U_{0,1}=0$ are evidently different from each other. This is due to the arbitrary phase difference between $m=0$ and $m=1$ emitted fields, and we argue it corresponds to the occurrence of the spontaneous breaking of the symmetry defined in Eq.~\ref{eq:phase_symmetry_m1}.\\
In fact, the stochastic nature of the perturbation coupled to the order parameter leads to the stabilization of a configuration randomly chosen among all the equivalent steady state configurations of the system. As a consequence, the features of the spontaneously broken symmetry group can be evidenced by measuring a suitable observable after independent realizations of the same equilibration process. This is supported by numerical results shown in Fig.~\ref{fig:3}. Here we report the behavior of the mean modulation over a sample of $N$ independent realizations of the steady state,i.e.,
 {\small 
 \begin{equation}\label{eq:mean_modulation_2}
   \Delta_{N}(x)=\frac{1}{N}\sum_{j=1}^{N}\Delta^{(j)}(x) \,,
 \end{equation}}
in which we define the modulation pattern of the $j$-th realization as 
\begin{equation}
    \Delta^{(j)}(x)=\lim_{t\to\infty}\left(\vert A^{(j)}_{out}\vert^2-\vert A^{(j)}_{0,out}\vert^2-\vert A^{(j)}_{1,out}\vert^2\right)/{\vert A^{(j)}_{0,out}\vert}.
\end{equation}
By construction, $\Delta^{(j)}(x)$ gives information about the interference pattern created by the superposition of $m=0$ and $m=1$ emissions. If the relative phase between such profiles is fixed and constant, all the modulation patterns $\Delta^{(j)}(x)$ will appear exactly the same for all realizations, and they will interfere constructively. As a consequence, we will observe the same $\Delta_{N}(x)$ for any $N$. On the contrary, if the different patterns correspond to independent realizations resulting from the spontaneous breaking a $U(1)$ symmetry, they will interfere destructively. This is a consequence of the arbitrarily chosen relative phase between the condensate and satellite states at finite $k_x$. In particular, we see that $\Delta_{N}(x)\to 0$ as $N \to \infty$ in this case, as shown in Fig.~\ref{fig:3}.

\begin{figure}[t]
    \centering
    \includegraphics[width=\textwidth]{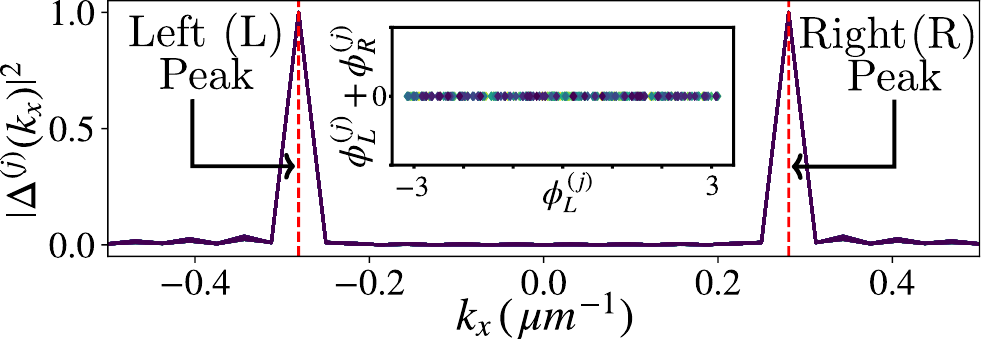}
    \caption{Spectral components, $\vert \Delta^{(j)}(k_x)\vert^2 $, associated to 300 ($j=1,\cdots, 300$) different steady state modulation profiles, $\Delta^{(j)}(x)$, as a function of $k_x$. The inset shows the sum of the Left and Right peaks phases ($\phi^{(j)}_{L}+\phi^{(j)}_{R}$) as a function of the Left peak phase, $\phi^{(j)}_{L}$, for each different simulation. Here we only assume $U_{0,1}= 0$, with $P_0\cdot \sigma=50$ ps$^{-1}$. Model parameters used in simulations are provided in the End Matter section.}
    \label{fig:4}
\end{figure}

We now report further quantitative analysis concerning data for $U_{0,1}=0$, and some further comments on the steady state features we observed in simulations that confirm the nature of the supersolid phase of polariton condensates in this scenario. First, we remind that all the modulation patterns $\Delta^{(j)}(x)$ obtained in this case have all an almost periodic behavior in space. To this purpose, we report in Fig.~\ref{fig:4} the absolute value of the Fourier components of $\Delta^{(j)}(x)$ as a function of $k_x$, calculated for the 300 different randomly generated profiles used for the previous analysis. In all cases, the spectrum appears strongly peaked around positive/negative $k_x$ values, corresponding to the difference between the condensate ($k_x=0$) and the side peaks on the $(1,\pm)$ bands located at $\pm\bar{k}_x\simeq\pm 0.3$ $\mu$m$^{-1}$, respectively. This is in agreement with the emission spectrum shown in Fig.~\ref{fig:2}d. In addition, the Fourier analysis gives us access to the phases $\phi^{(j)}_{L}$ and $\phi^{(j)}_{R}$ of the Left ($k_x\simeq -\bar{k}_x$) and Right ($k_x\simeq \bar{k}_x$) propagating components, respectively. As shown in the inset of Fig.~\ref{fig:4}, we evidently get $\phi^{(j)}_{L}+\phi^{(j)}_{R}=0$ for any $j\in [1,300]$. 
In addition, numerical results show that $\phi^{(j)}_{L}$ takes values randomly spanning the whole $[-\pi,\,\pi]$ interval. Therefore,  outside the region where emission is dominated by the dark nature of the condensate (i.e., around $x=0$, as visible in Fig.~\ref{fig:3}b, $N=1$), the main contribution to the spatial modulation $\Delta^{(j)}(x)$ behaves as $\cos ( \bar{k}_x \cdot x+ \phi^{(j)}_{L})$, with $\bar{k}_x\simeq 0.3 $ $\mu$m$^{-1}$. These results unequivocally show the spontaneous breaking of a phase symmetry in real space, which gives us the ultimate definition of the polariton supersolid state. 

As a final comment, we notice that modulations have a spatial periodicity that is compatible with $2\pi/\bar{k}_x\simeq 21$ $\mu$m, totally unrelated from the 1DPC lattice period, typically set as $a\sim 0.25$ $\mu$m in these systems~\cite{Riminucci2023prl}. Moreover, the spatial periodicity of this polariton supersolid is determined by the mode dispersion of the $m=1$ branches, through the wavevector of the isoenergetic states that are resonant with the energy of the condensate. Since the latter can be controlled through the input power (i.e., through the blue shift of the condensate), the modulations periodicity can be controlled as well (see the schematic illustration in Fig.~S3 of the SM file). This peculiar aspect differentiates this non-equilibrium, driven-dissipative supersolid from realizations reported in ultracold atomic gases, in which the periodicity is solely determined by the interactions terms.\\ 

\section{Summary} 
We have proposed a realistic configuration in which a new phase of matter simultaneously breaking the phase and the translational symmetries can be observed in the stationary regime of a condensate of exciton-polaritons. We define this phase as the polariton supersolid. Besides establishing the theoretical aspects of this novel phase of quantum fluids of light, we have explicitly pointed out experimentally accessible signatures of the simultaneous breaking of the condensate phase and its spatial symmetries. In emission, we have identified the presence of a double threshold in the emission from the polariton side-branches as the fingerprint of the polariton supersolid phase transition on top of the polariton condensate, in addition to the onset of a spatial modulation. Besides providing a further platform where such exotic states of matter can be studied, polariton supersolids intrinsically differentiate from their atomic counterparts owing to their driven-dissipative nature. As for the standard non-equilibrium condensation \cite{claude2023observation}, this is anticipated to give peculiar dispersion laws for the two Goldstone modes associated to the spontaneously broken symmetries \cite{Tanzi2019,Guo2019,Natale2019prl,LeonardScience2017,Geier2021prl}, with potential consequences 
% onto the superfluid properties \cite{PhysRevLett.105.020602} and 
on the long-distance coherence of the corresponding order parameters \cite{Fontaine_Nature_KPZ,Daviet2024prl}.
\DN{Given the subtlety of the concept of superfluidity in non-equilibrium systems \cite{PhysRevLett.105.020602}, in particular for condensates in gap states \cite{Grudinina2023}, ongoing work is being devoted to assess the superfluid properties of this new state of matter. In this respect, our theoretical investigation (see e.g. the SM file, "Instability analysis") reveals that modulations spontaneously created on top of the polariton condensate are at rest, so unconditionally subsonic and perfectly compatible with superfluidity.} 
% , also taking into account the persistence of metastable supercurrents that only depend on the long-range superfluid coherence and not on the Bogoliubov dispersion \cite{PhysRevLett.105.020602}.\\

\textbf{Acknowledgments. } 
This project was partly funded by the European Union–Next Generation EU within the Italian Ministry of Research (MUR) PNRR project PE0000023-NQSTI.
I.C. acknowledges financial support from the Provincia Autonoma di Trento, from the Q@TN initiative. The authors acknowledge useful scientific discussions with G.I. Martone. D.N. warmly acknowledges A. Tripodo  for helpful discussions.
\section{End Matter to:\\ Supersolidity of polariton condensates in photonic crystal waveguides}
\subsection{Theoretical model} The model considered in this Letter is a generalization of a previously introduced system of coupled equations for the coupled exciton-photon amplitudes in a multi-component formulation~\cite{Nigro2023prb}, which explicitly reads:
\begin{subequations}
\label{em_allequations} 
\begin{eqnarray}
\dot{n}_I&=&-\gamma_{I}n_I + \gamma_{A\to I}n_A + P,\label{em_eq:inactive_reservoir_eom}\\
\dot{n}_A&=&-\gamma_{A}n_A + \gamma_{I\to A}n_I -g n_{A}\langle \vec{\psi},G \vec{\psi} \rangle ,\label{em_eq:active_reservoir_eom}\\
 \dot{\vec{\psi}}&=&-\frac{i}{\hbar}\left(H\vec{\psi}+g_{xx}\vec{W}[\vec{\psi}]\right)+\left(\frac{g}{2}n_A G -\Gamma\right)\vec{\psi}+\vec{\xi},\quad \label{em_eq:psi_eom}
\end{eqnarray}
\end{subequations}
in which $n_I\equiv n_I(x,t)$ and $n_A\equiv n_A(x,t)$ are the inactive and active exciton reservoir densities, respectively, $P=P_0\exp{\{-(x^2)/(2\sigma^2)\}}$ is the spatially-dependent incoherent pump in continuous wave (cw), and $\vec{\xi}=\vec{\xi}(x,t)$ denotes the random fluctuating noise vector. In this picture, the non-resonant cw source injects particles into the inactive reservoir (i.e., high-energy excitations). The active reservoir density, instead, accounts for excitations satisfying the proper energy-momentum conditions to be converted into polaritons. 
In order to allow for population exchange between the reservoirs, and to account for particle losses, two different gain terms and two particle decay channels are included, respectively. In particular, the former are described by $\gamma_{A\to I}$ and $\gamma_{I\to A}$, while the latter by the rates $\gamma_{I}$ and $\gamma_{A}$. 
Particle transfer from $n_A$ to polariton excitations is made possible by the presence of the two terms proportional to $g$: the one in Eq.~\ref{em_eq:active_reservoir_eom} is responsible for the depletion of the $n_A$ density, whenever some particle population is stored into $\vec{\psi}$, with $\langle \vec{\psi},\vec{\psi}^{\prime}\rangle= \sum_j \psi^{*}_j \psi^{\prime}_j$ denoting the standard hermitian scalar product.
In particular, we assume the following ordering for photonic and excitonic components in our multi-component vector:
{\small
\begin{equation}
\vec{\psi}\,^{T}=(A_{0,+},X_{0,+},A_{0,-},X_{0,-},A_{1,+}, X_{1,+},A_{1,-},X_{1,-}) \, ,
\end{equation}}
in which $O^{T}$ denotes the transposed of $O$. Particles scattered by the active reservoir are fed into the polariton subsystem by the gain term in Eq.~\ref{em_eq:psi_eom}. In particular, the operator $G$ in Eqs.~\ref{em_eq:active_reservoir_eom},\ref{em_eq:psi_eom} is a $8\times 8$ matrix satisfying the constraints already discussed in Ref.~\cite{Nigro2023prb}. Specifically, we assume the following form for $G$:
\begin{equation}
{\footnotesize
    G=\frac{1}{4}
\begin{pmatrix}
     1-\alpha& 0 &0 &0 &0 &0 &0 &0 \\
     0& \alpha& 0 &0 &0 &0 &0 &0 \\
     0& 0 & 1-\alpha& 0 &0 &0 &0 &0  \\
     0 &0 &0 & \alpha& 0 &0 &0 &0 \\
     0 &0 &  0& 0 & 1-\alpha& 0 &0 &0   \\
     0& 0 &0 &  0& 0 & \alpha& 0 &0   \\
     0& 0& 0 &0 &  0& 0 & 1-\alpha& 0    \\
     0& 0& 0& 0 &0 &  0& 0 & \alpha    \\
\end{pmatrix},
}
\end{equation}
with $\alpha$ being a phenomenological parameter.\\
We notice that the scattering from the active reservoir to $\vec{\psi}$ is triggered by the noise vector, $\vec{\xi}$, here assumed to be a complex gaussian-correlated noise (see SM file for further details). However, such a perturbation does not necessarily lead to an instability of the vacuum configuration, $\vec{\psi}=\vec{0}$. Indeed, in order to observe a stationary macroscopic population in polariton states encoded into $\vec{\psi}$, the gain rate (controlled by $\frac{g}{2}n_A G$) must overcome losses (controlled by $\Gamma$) of the underlying eigenmodes. Such states are determined by the (i) band structure ($H_{0}(x)$) and (ii) the presence of a positive local energy blue-shift ($V(x,t)$) around normal incidence. In other words, condensation and physics beyond condensation are ultimately determined by the competition of $H(x,t)=H_{0}(x) + V(x,t)$, gain terms, and losses. In particular, we used the following representations:
\begin{equation}\label{em_eq:Mat_H_0}
{\footnotesize
    \frac{H_{0}}{\hbar}=
\begin{pmatrix}
    \omega_{0+}& \Omega_{R0} &  U_{0,0} & 0 & 0 & 0 & U_{0,1} & 0 \\
    \Omega_{R0} & \omega_X & 0 & 0 & 0 & 0 & 0 & 0 \\
     U_{0,0} & 0 & \omega_{0-}& \Omega_{R0} & U_{0,1} & 0 & 0 & 0   \\
    0 & 0 &  \Omega_{R0} & \omega_X & 0 & 0 & 0 & 0  \\
    0 & 0 & U_{0,1} & 0 & \omega_{1+}& \Omega_{R1} & 0 & 0\\
    0 & 0 & 0 & 0 & \Omega_{R1} & \omega_X & 0 & 0\\
     U_{0,1} & 0 &0 & 0 & 0 & 0 & \omega_{1-}&  \Omega_{R1} \\
    0 & 0 &0 & 0 & 0 & 0 & \Omega_{R1} & \omega_X \\
\end{pmatrix}
    }
\end{equation}
and 
\begin{equation}\label{em_eq:Potential}
    V(x,t)=\frac{1}{2}(w_I n_I(x,t)+w_A n_A(x,t)) \mathbb{1}_8 \, ,
\end{equation}
in which $\mathbb{1}_8$ denotes the $8\times 8$ identity matrix.
In Eq.~\ref{em_eq:Mat_H_0} $\omega_{m,l}\equiv E_{m,l}(x)/\hbar$ are $x$-dependent quantities, and their specific form follows from $E_{m,\pm}(x)=\hbar \omega_m \mp i \hbar v_{g,m} \partial_x$. These quantities describe the energy-wavevector dispersions of the photonic modes, labelled $A_{m,l}$. In particular, for a given $m$, $\hbar \omega_m$ gives the energy at which $l=+$ and $l=-$ dispersions cross, while $\pm v_{g,m}$ give the slopes of the $l=\pm$ component. The nearly flat dispersion of excitons is encoded into $\omega_{X}\equiv E_{X}/\hbar$. Linear coupling between the photonic mode $A_{m,l}$ and the exciton replica $X_{m,l}$ is controlled by the Rabi term, $\Omega_{R,m}$. Linear coupling between the photonic modes, $(m,+)$ and $(m',-)$, quantified by matrix elements $U_{m,m'}$, accounts for the possible interactions induced by the 1DPC potential. In the analysis reported in our Letter, we assumed (i) $U_{0,0}\neq 0$, (ii) $U_{1,1}=0$, and (iii) we compared results obtained in the absence ($U_{0,1}= 0$) or presence ($\hbar U_{0,1}=1 meV$) of coupling between $m=0$ and $m=1$ branches.\\

\begin{figure}[ht]
    \centering
    \includegraphics[width=\linewidth]{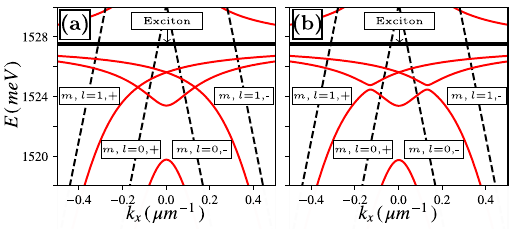}
    \caption{Example of polariton band dispersion (red solid lines) obtained by diagonalization of $H_{0}$, as a function of horizontal wave vector, $k_x$ (see main text) for \textbf{(a)} $U_{0,1}=0$ and \textbf{(b)} $U_{0,1}=1\,meV$. Dashed lines describe the energy-wavevctor dispersion of bare photonic modes $A_{m,l}$. The thick horizontal line corresponds to the bare exciton resonance.}
    \label{em_fig:example_bands}
\end{figure}

The polaritonic band structure associated to $H_{0}(x)$ is obtained by substituting $-i \partial_{x} $ with $k_x$, and by diagonalizing the resulting $8\times 8$ matrix (thus giving 8 polariton bands). Out of these 8 modes, only polariton bands below the exciton resonance are relevant to our purposes. In particular, the bands obtained in the two cases discussed in our Letter ($U_{0,1}= 0$ or $U_{0,1}\neq 0$) are shown in Fig.~\ref{em_fig:example_bands}. A crucial point is the presence of a sizeable energy-gap at $k_x=0$. Under non-resonant pumping, polaritons are observed to condense in flat modes appearing within this gap. This phenomenon has been already discussed, and attributed to the presence of a local positive blue-shift term induced by the reservoir populations, which traps negative effective mass polaritons~\cite{Nigro2023prb}. In our model, terms accounting for such a space-dependent blue-shift are included into the effective repulsive potential, $V(x,t)$, as given in Eq.~\ref{em_eq:Potential}. In particular, $w_{I}$ and $w_{A}$ are two phenomenological constants, which  weigh the contributions associated to $n_I$ and $n_A$, respectively.\\

We now pay attention to parameters entering into the $\Gamma$ operator, whose form reads:
\begin{equation}\label{em_eq:Gamma_matrix}
{\small 
    \Gamma=
\begin{pmatrix}
    \gamma_{ph,0}& 0 & \gamma_{int,0} & 0 & 0 & 0 & 0 &0 \\
    0 & \gamma_X & 0 & 0 & 0 & 0 & 0 & 0 \\
    \gamma_{int,0} & 0 & \gamma_{ph,0}& 0 & 0 & 0 & 0 & 0   \\
    0 & 0 & 0 & \gamma_X & 0 & 0 & 0 & 0  \\
    0 & 0 & 0 & 0 & \gamma_{ph,1}& 0 & 0 & 0\\
    0 & 0 & 0 & 0 &0 & \gamma_X & 0 & 0\\
    0 & 0 &0 & 0 & 0 & 0 & \gamma_{ph,1}& 0 \\
    0 & 0 &0 & 0 & 0 & 0 &0 & \gamma_X  \\
\end{pmatrix}
}
\end{equation}
Terms on the diagonal control the lifetimes of bare modes encoded into the components of $\vec{\psi}$. Specifically, photonic and excitonic losses are controlled through $\gamma_{ph,m}$ and $\gamma_X$, respectively. The off-diagonal terms, $\gamma_{int,0}$, are associated to the fact the underlying $m=0$ photonic branch supports a symmetry-protected bound-state in the continuum at $k_x=0$. In particular, one must have  $\gamma_{ph,0} \geq \vert \gamma_{int,0} \vert $ as pointed out in Ref.~\cite{Nigro2023prb}. \\

We conclude this paragraph by showing the explicit form of the term $\vec{W}[\vec{\psi}]$. As discussed in the main text, such a term can be regarded as the functional derivative, $\delta \mathcal{U}/\delta \vec{\psi}^{*}$, of a contact potential $\mathcal{U}$, that is
\begin{equation}\label{em_eq:nonlinear_coupling}
\mathcal{U}=\int dx \left(X^*_{0,+}X^*_{0,-}X_{1,+}X_{1,-} + c.c. \right),
 \end{equation}\\
 which accounts for energy contributions associated to the conversion of a couple $(X_{1,+},X_{1,-})$ to $(X_{0,+},X_{0,-})$, and vice versa. Since $\mathcal{U}$ does not depend on photonic components, $\delta \mathcal{U}/\delta A^{*}_{m,l}=0$. On the contrary, the functional derivative of $\mathcal{U}$ with respect to any excitonic component is non zero. Specifically, one obtains that 
 \begin{equation}
     \frac{\delta}{\delta X^{*}_{m,\pm}}\mathcal{U} = X^{*}_{m,\mp}X_{m^{'},+}X_{m^{'},-} \,\,(m\neq m^{'})
 \end{equation}
 As a consequence, the explicit form of $\vec{W}[\vec{\psi}]$ reads:
  {\small
 \begin{equation}
     \vec{W}^{T}[\vec{\psi}]=\left(0,\frac{\delta \mathcal{U}}{\delta X^{*}_{0,+}},0,\frac{\delta \mathcal{U}}{\delta X^{*}_{0,-}},0,\frac{\delta \mathcal{U}}{\delta X^{*}_{1,+}},0,\frac{\delta \mathcal{U}}{\delta X^{*}_{0,-}}\right)
 \end{equation}}

\subsection{Simulation parameters}. We conclude this End Matter section by listing the values of parameters used in the numerical simulations whose results are shown in the text:
$\hbar \omega_0$=1530.2  $meV$; $\hbar \omega_1$=1550.2 $meV$;  $v_{g,0}$ =110.4 $\mu m /ps$; $v_{g,1}$ = 110.4 $\mu m /ps$; $E_X$ = 1527.5 $meV$; $\hbar \Omega_{R,0}$ = 6.8 $meV$; $\hbar \Omega_{R,1}$ = 6.8 $meV$; $\hbar U_{0,0}$ = 4.45 $meV$;  $\hbar U_{0,1}$ =1 or 0 $meV$; $\hbar\gamma_{ph,0}$ = 0.1 $meV$; $\hbar\gamma_{int,0}$ = 0.1 $meV$; $\hbar\gamma_{ph,1}$ = 0.1 $meV$; $\hbar \gamma_X$ = 0.01 $meV$; $\hbar \gamma_{A}$= 0.1 $\hbar \gamma_X$; $\hbar \gamma_{I}$ = 0.2 $\hbar \gamma_X$; $\hbar \gamma_{I\to A}$= 0.05 $\hbar \gamma_X$; $\hbar \gamma_{A\to I}$= 0.05 $\hbar \gamma_X$; $g$ = $3\times 10^{-2} meV \mu m$; $\alpha=10^{-3}$; $g_{xx}$ = 1 $meV\, \mu m$; $w_{A}=0$;  $w_I $ = $3.3\times 10^{-3} meV \mu m$.\\

Further details concerning the model, the choice of parameter values, and comments about the physical content of model in Eq.~\ref{em_allequations} are reported in the SM section, as well as the numerical scheme employed in our analysis.
\section{Supplementary Material to:\\ {Supersolid state of polariton condensates in photonic crystal waveguides}}

\subsection{Summary}
In this document we provide further details concerning the theoretical framework and the results reported in the main text of this communication. In Sec.~\ref{sm_sec:ops} we provide further information about the theoretical model reported in the End Matter section, as well as comments about the choice of values of parameters employed in the numerical simulations to obtain the results shown in the main text. In Sec.~\ref{sm_sec:additional_numerical_results} we discuss further numerical results, that are complementary to those shown in the main text. A separate section of this Supplementary Material file (Sec.~\ref{sm_sec:scaling}) is devoted to a theoretical discussion on the scaling behavior of the dynamical equations. In particular, their steady state solutions will be analyzed in terms of rescaled parameters, showing that different nonlinear regimes (in terms of parameters) are actually equivalent upon proper rescaling of the model parameters. Subsequently, in Sec.~\ref{sm_sec:instability} we briefly comment on the emergence of the second emission threshold analyzed in the main text in the absence of linear coupling, that is $U_{0,1}=0$. In particular, we report an instability analysis providing an explicit relation between the exciton-exciton nonlinearity, the condensate profile in wave vector space, and the loss rates of the relevant polariton branches. The form of the equations of motion makes also manifest the phase freedom spontaneously broken at the onset of the supersolid phase, which is the main claim reported in the manuscript. Finally, for the sake of completeness, in Sec.~\ref{sm_sec:numerical_details} we provide details concerning the numerical scheme employed to determine the steady-state configuration, and the random noise used to initialize the numerical simulations.  

\subsection{Operators and relevant parameters}\label{sm_sec:ops}
For the sake of completeness, we hereby report  the model already defined in the main text, by focusing first on the part concerning the polariton evolution, and subsequently on the interaction term between $\vec{\psi}$ and the active reservoir, $n_A$. The full system of coupled differential equations reads
\begin{equation}\label{sm_eq:inactive_reservoir_eom}
   \dot{n}_I=-\gamma_{I}n_I + \gamma_{A\to I}n_A + P(x),
\end{equation}
\begin{equation}\label{sm_eq:active_reservoir_eom}
    \dot{n}_A=-\gamma_{A}n_A + \gamma_{I\to A}n_I -g n_{A}\langle \vec{\psi},G \vec{\psi} \rangle
\end{equation}
\begin{equation}\label{sm_eq:psi_eom}
   \dot{\vec{\psi}}=-\frac{i}{\hbar}H(x,t)\vec{\psi}- \Gamma \vec{\psi}+\frac{g}{2}n_A G \vec{\psi}-\frac{i}{\hbar}g_{xx}\vec{W}[\vec{\psi}]+\vec{\xi} \, ,
\end{equation}
where the main quantities, i.e., parameters and variables, have been defined in the main text already. As it is mentioned in the manuscript, the entries $\vec{\psi}$ correspond to the wavefunctions of the relevant photonic modes, $\{A_{m,l}\}$, as well as the associated exciton fields, $\{X_{m,l}\}$ (with index labels $m=0,1$ and $l=+,-$, respectively), which have been used to formulate our multi-band theory in a general framework. More explicitly, we hereby report the multi-component vector $\vec{\psi}$ as
\begin{equation}\label{sm_eq:psi_def}
    \vec{\psi}(x,t)=\left(
\begin{array}{c}
     A_{0,+}(x,t)\\
     X_{0,+}(x,t)\\
     A_{0,-}(x,t)\\
     X_{0,-}(x,t)\\
     A_{1,+}(x,t)\\
     X_{1,+}(x,t)\\
     A_{1,-}(x,t)\\
     X_{1,-}(x,t)\\
\end{array}
    \right)
\end{equation}
in which $x$ denotes the position along the one-dimensional photonic crystal (1DPC) lattice, and $t$ is the time.\\
With such an ordering of modes, the operators $H_0(x)$ and $V(x,t)$ defining the total Hamiltonian, $H(x,t)=H_{0}(x) + V(x,t)$, are represented in the following matrix forms
\begin{equation}\label{sm_eq:Mat_H_0}
    H_{0}(x)=\left(
\begin{array}{cccc|cc|cc}
    E_{0,+}(x)& \hbar \Omega_{R,0} & \hbar U_{0,0} & 0 & 0 & 0 & \hbar U_{0,1} & 0 \\
    \hbar \Omega_{R,0} & E_X & 0 & 0 & 0 & 0 & 0 & 0 \\
    \hbar U_{0,0} & 0 & E_{0,-}(x)& \hbar \Omega_{R,0} & \hbar U_{0,1} & 0 & 0 & 0   \\
    0 & 0 & \hbar \Omega_{R,0} & E_X& 0 & 0 & 0 & 0  \\
    \hline
    0 & 0 & \hbar U_{0,1} & 0 & E_{1,+}(x)& \hbar \Omega_{R,1} & 0 & 0\\
    0 & 0 & 0 & 0 &\hbar \Omega_{R,1} & E_X & 0 & 0\\
    \hline
    \hbar U_{0,1} & 0 &0 & 0 & 0 & 0 & E_{1,-}(x)& \hbar \Omega_{R,1} \\
    0 & 0 &0 & 0 & 0 & 0 &\hbar \Omega_{R,1} & E_X \\
\end{array}
    \right)
\end{equation}
and 
\begin{equation}\label{sm_eq:Potential}
    V(x,t)=\frac{1}{2}\mathbb{1}_8(w_I n_I(x,t)+w_A n_A(x,t)) \, ,
\end{equation}
in which $\mathbb{1}_8$ is the $8\times 8$ identity.
Each of the bare photonic bands that are associated to the folding within the first Brillouin zone of the periodic lattice are represented in the vicinity of $k_x=0$, assuming a linear dispersion for each of the $m$-th waveguide modes in the basis. The latter is explicitly given as $E_{m,\pm}(x)=\hbar \omega_m \mp i \hbar v_{g,m} \partial_x$ for each value of $m$ (with equal phase and group velocities, given as $v_{g,m}$). Here, $\hbar \omega_m$ denotes the energy at which the propagating ($l=+$) and counter-propagating ($l=-$) components of a given $m$-th photonic branch cross at $k_x=0$. 
%The parameter $v_{g,m}$ corresponds to the (absolute value of the) group-velocity of the $m$-th branch. 
In this straightforward representation, propagating and counter-propagating photonic eigenmodes of both $m=0$ and $m=1$ branches are linearly coupled to exciton eigenmodes at fixed energy $E_X$ (here assumed to be dispersionless on the relevant wave vector scale, owing to the large exciton effective mass) via the off-diagonal Rabi couplings, here denoted as $\hbar \Omega_{R,m}$. Furthermore, we assume that the 1DPC lattice, with its periodic modulation of the refractive index, introduces a significant linear coupling between propagating and counter-propagating $m=0$ components, respectively, quantified by the off-diagonal energy term $\hbar U_{0,0}$ in Eq.~\ref{sm_eq:Mat_H_0}. As already mentioned in the main text, we can also assume a linear coupling between $m=0$ and $m=1$ components, quantified by the term $\hbar U_{0,1}$. The coupling between  propagating and counter-propagating $m=1$ components is set to zero, instead. Notice that we can also set $U_{0,1}=0$, which can always be fulfilled by proper device engineering, e.g., through the filling fraction or the etching depth of the photonic lattice \cite{GeracePRE2004}. The values of parameters entering in $H_{0}(x)$ and used in our calculations are reported in Tab.~\ref{sm_tab:Hamiltonian_parameters}, for completeness. They are compatible with parameters obtained in actual samples fabricated in III-V semiconductor technology, e.g., from Ref.~\cite{Riminucci2023prl}. Of course, the whole theoretical framework reported in this work can be adapted to different material platforms, when needed. 

\begin{table}[]
    \centering
    \begin{tabular}{|c|c|}
         \hline
         $\hbar \omega_0$ & 1530.2 $meV$ \\
         \hline
         $\hbar \omega_1$ & 1550.2 $meV$ \\
         \hline
         $v_{g,0}$ & 110.4 $\mu m /ps$\\
         \hline
         $v_{g,1}$ & 110.4 $\mu m /ps$\\
         \hline
         $E_X$ & 1527.5 $meV$\\
         \hline
         $\hbar \Omega_{R,0}$ & 6.8 $meV$\\
         \hline
         $\hbar \Omega_{R,1}$ & 6.8 $meV$\\
         \hline
         $\hbar U_{0,0}$ & 4.45 $meV$\\
         \hline
         $\hbar U_{0,1}$ & 1 $meV$ or 0 $meV$\\
         \hline
    \end{tabular}
    \caption{Parameter values assumed in this work, and employed in simulations to produce the results reported in the main text. These parameters enter into the definition of the Hamiltonian model, $H_0(x)$, represented in Eq.~\ref{sm_eq:Mat_H_0}.}
    \label{sm_tab:Hamiltonian_parameters}
\end{table}

In recent experiments~\cite{Ardizzone_2022,Riminucci2023prl}, it has been shown that the presence of an external non-resonant driving laser leads to the formation of quantized defect modes within the polaritonic gap opened around $k_x=0$ by the $\hbar U_{0,0}$ term in $H_0$. In particular, they appear as dispersionless energy resonances in emission (i.e. flat in energy), while their spread in $k_x$ is compatible with the average size of the real space reservoir distribution, reported in momentum (Fourier) space. For such a reason, we assume the local energy blueshift to affect both photonic and excitonic modes, resulting in a potential term proportional to the reservoir densities. Physically speaking, the two effects can be reduced to a refractive index change on one side, and electronic band-gap renormalization on the other, due to free-carrier populations in the conduction and valence bands of the underlying semiconducting materials. In our model, we simply assume each excitonic and photonic mode to be equally affected by such a background density, resulting into the expression for $V(x,t)$ that is reported in Eq.~\ref{sm_eq:Potential}. We stress that this simplifying assumption has no practical consequences on the results we are focusing in the following. In particular, for the results reported in the main text we have assumed $V(x,t)\propto n_I(x,t)$ (i.e., $w_A=0$), while $w_I$ has been chosen in order to obtain the appearance of a single defect mode within the energy gap formed by $m=0$ polaritons around $k_x=0$, within the whole range of injection rates considered. In the simulations, we have specifically set $w_I=3.3\times 10^{-3} meV \cdot \mu m$.

The loss matrix defined in the model of coupled equations  explicitly reads
\begin{equation}\label{sm_eq:Gamma_matrix}
    \Gamma=\left(
\begin{array}{cccc|cc|cc}
    \gamma_{ph,0}& 0 & \gamma_{int,0} & 0 & 0 & 0 & 0 &0 \\
    0 & \gamma_X & 0 & 0 & 0 & 0 & 0 & 0 \\
    \gamma_{int,0} & 0 & \gamma_{ph,0}& 0 & 0 & 0 & 0 & 0   \\
    0 & 0 & 0 & \gamma_X & 0 & 0 & 0 & 0  \\
    \hline
    0 & 0 & 0 & 0 & \gamma_{ph,1}& 0 & 0 & 0\\
    0 & 0 & 0 & 0 &0 & \gamma_X & 0 & 0\\
    \hline
    0 & 0 &0 & 0 & 0 & 0 & \gamma_{ph,1}& 0 \\
    0 & 0 &0 & 0 & 0 & 0 &0 & \gamma_X  \\
\end{array}
    \right)
\end{equation}
in which the top left $4\times 4$ block accounts for losses and the dissipative linear coupling within the $m=0$ modes sub-sector. In particular, while diagonal terms account for intrinsic loss rates of both exciton and photon modes labelled with either $m=0$ or $m=1$, the $\gamma_{int,0}$ parameter describes interaction through the continuum of modes above the light cone mediated by the periodic photonic crystal pattern. These terms derive from the symmetry properties of the underlying photonic eigenmodes \cite{GeracePRE2004}, and are inherited by the gap-confined eigenstates created by the local blueshift, as nicely detailed in Ref.~\cite{SigurdssonNguyenNguyen2024}. On the other hand, since the bare photonic branch crossing between $m=1$ components occurs at energy well above $E_X$ in the present analysis, we assume such non-diagonal loss terms to be negligible ($U_{1,1}=0$). Therefore, the loss matrix in the $m=1$ sector is assumed to be purely diagonal, as given in Eq.~\ref{sm_eq:Gamma_matrix}. Specifically, the parameter values of these loss rates, given as full width at half maximum (i.e., linewidths) of energy resonances ($\hbar \gamma_{j}$) are reported in Tab.~\ref{sm_tab:Gamma_parameters}. These values, used in our simulations, are compatible with actual parameters extracted from experimental data in the reference configurations for the present work \cite{Ardizzone_2022,Riminucci2023prl}.

After fixing the exciton linewidth, $\hbar \gamma_X$, the other parameters used to capture the reservoir dynamics are assumed as follows: $\hbar \gamma_{A}=0.1 \hbar \gamma_X$; $\hbar \gamma_{I}=0.2 \hbar \gamma_X$; $\hbar \gamma_{I\to A}=0.05 \hbar \gamma_X$; $\hbar \gamma_{A\to I}=0.05 \hbar \gamma_X$. We stress that such parameters are \emph{phenomenological} quantities and their values, together with the parameters $w_A$ and the parameter $g_{xx}$ (see the discussion in Sec.~\ref{sm_sec:scaling}), can always be redefined to match the particular experimental regime of interest (e.g., the condensation threshold, or the energy of gap-confined modes as a function of the injection rate, associated to the pump strength and spatial profile, $P(x)$). Thus, if on one hand different choices for the specific parameter values are always possible, on the other it is reasonable to assume that reservoir exciton states should have lifetimes compatible with or even larger than the bare quantum well excitons. This conjecture follows from time-resolved experimental data reported for devices of this type and performed in pulsed excitation regime~\cite{Ardizzone_2022}. Indeed, by assuming that the excitations reservoir is responsible for the local blueshift of the lower polariton branch, thus trapping negative-mass states within the energy gap, it is then reasonable to assume that the reservoir particles live longer than (or at least as long as) polaritons relaxed in such an energy region. In particular, since the lifetime of such polariton states is bound by the bare exciton lifetime, $\gamma_X^{-1}$, the loss rates for the reservoir states should satisfy the following constraint: $\gamma_{A/I}\leq \gamma_{X}$. In addition, we further notice that very similar parameters have been used to compare with available (and unpublished) experimental data, which further confirms our conjecture.

\begin{table}[]
    \centering
    \begin{tabular}{|c|c|}
    \hline
        $\hbar\gamma_{ph,0}$ & 0.1 $meV$ \\
        \hline
        $\hbar\gamma_{int,0}$ & 0.1 $meV$\\
        \hline
        $\hbar \gamma_X$ & 0.01 $meV$\\ 
        \hline
        $\hbar\gamma_{ph,1}$ & 0.1 $meV$ \\
        \hline
    \end{tabular}
    \caption{Energy linewidths (i.e., full width at half maximum of energy resonances) for the relevant bare photonic and excitonic eigenmodes, as used in numerical simulations and appearing in the definition of the $\Gamma$ matrix represented in Eq.~\ref{sm_eq:Gamma_matrix}.}
    \label{sm_tab:Gamma_parameters}
\end{table}

We now discuss the two terms in the equations of motion that couple the evolution of the active reservoir density, $n_A$, to $\vec{\psi}$. The first term, which induces an extra decay contribution to the active reservoir whenever a macroscopic population is stored into the photonic and excitonic modes, explicitly reads
\begin{equation}\label{sm_eq:extra_loss_term_active_reservoir}
    M_{loss}=-g n_A \langle \vec{\psi},G \vec{\psi}\rangle = -g\,n_A \sum_{i,j}\psi^{*}_i G_{i,j}\psi_{j} \, ,
\end{equation}
in which the indices $i$ and $j$ run over the 8 labels associated to the $\{A_{m,l}\}$ and the $\{X_{m,l}\}$ modes, respectively. Following the detailed discussion reported in Ref.~\cite{Nigro2023prb}, the matrix $G$ has to satisfy the following constraints:
\begin{equation}
    G^{\dagger}=G\quad \mbox{(hermitian)},\,G\geq 0\,\mbox{(non-negative)},\,\mbox{Tr}[G]=1\,\mbox{(unit\,trace)} \, ,
\end{equation}
which guarantees particle conservation (i.e., processes for which a single reservoir particle may only create a single polariton excitation). 
In the present case, we assume all the photonic modes to be equally coupled to the reservoir. In addition, even though other parametrizations are possible, for the sake of simplicity we assume $G$ to be diagonal and independent on $k_x$, which gives the following representation for the coupling matrix
\begin{equation}
    G=\frac{1}{4}\left(
\begin{array}{cccc|cc|cc}
     1-\alpha& 0 &0 &0 &0 &0 &0 &0 \\
     0& \alpha& 0 &0 &0 &0 &0 &0 \\
     0& 0 & 1-\alpha& 0 &0 &0 &0 &0  \\
     0 &0 &0 & \alpha& 0 &0 &0 &0 \\
     \hline
     0 &0 &  0& 0 & 1-\alpha& 0 &0 &0   \\
     0& 0 &0 &  0& 0 & \alpha& 0 &0   \\
     \hline
     0& 0& 0 &0 &  0& 0 & 1-\alpha& 0    \\
     0& 0& 0& 0 &0 &  0& 0 & \alpha    \\
\end{array}
    \right)
\end{equation}
in which $\alpha$ is a parameter such that $0\leq \alpha\leq 1$, which controls the relative coupling to the active reservoir between photonic ($1-\alpha$) and excitonic ($\alpha$) eigenmodes. Notice that particles lost by the active reservoir are directly fed into the polariton subsystem via the gain term appearing in Eq.~\ref{sm_eq:psi_eom}, which is written as
\begin{equation}
    M_{gain}=\frac{1}{2}g\,n_{A}G\vec{\psi} \, .
\end{equation}
Specifically, the numerical simulations reported in the main text have been obtained by setting $g=3\times 10^{-2} meV \mu m$ and $\alpha=10^{-3}$.

We now consider the nonlinear term that in our framework is responsible for exciton-exciton scattering. This interaction contribution is explicitly included in the model via the term $g_{xx}\vec{W}[\vec{\psi}]$. The quantity $\vec{W}[\vec{\psi}]$ is a vector having the same dimension as $\vec{\psi}$. In this context, by using the same ordering of components provided in Eq.~\ref{sm_eq:psi_def}, the component $j$ of $\vec{W}[\vec{\psi}]$, that is $\vec{W}_{j}$ is given by
\begin{equation}\label{sm_eq:definition_nonlinearity}
\vec{W}_{j}=\frac{\delta}{\delta \psi^{*}_{j}}\mathcal{U}[\vec{\psi}] \quad \mbox{with}\quad\mathcal{U}[\vec{\psi}]=\int dx [X^*_{0,+}X^*_{0,-}X_{1,+}X_{1,-} + X_{0,+}X_{0,-}X^*_{1,+}X^*_{1,-}  ],
\end{equation}
with $X_{m,l}\equiv X_{m,l}(x,t)$.
In particular, in this context the functional derivative is understood as follows
\begin{equation}
    \frac{\delta}{\delta \psi^{*}_{j}}\mathcal{U}[\vec{\psi}]=\lim_{\epsilon \to 0} \frac{\mathcal{U}[\vec{\psi}+\delta\vec{\psi}]-\mathcal{U}[\vec{\psi}]}{\epsilon},
\end{equation}
where the component $i$ of the variation $\delta\vec{\psi}$ is given by
\begin{equation}
    \delta\vec{\psi}_{i}=\left\{
    \begin{array}{cc}
        0 & \mbox{if}\,i \neq j \\
        \epsilon \delta(x-x^{'}) & \mbox{if}\,i =j
    \end{array}
    \right.
\end{equation}
with $\delta(x)$ being the Dirac delta distribution. Since $\mathcal{U}[\vec{\psi}]$ does not depend on photonic modes, functional derivatives of $\mathcal{U}$ in the directions of photonic modes is identically zero, that is
\begin{equation}
     \frac{\delta}{\delta A^{*}_{m,l}}\mathcal{U}[\vec{\psi}]=0, \quad m=0,1\,\,\mbox{and}\,\,l=+,-
\end{equation}
On the contrary, functional derivatives with respect to exciton modes are non zero. For the sake of clarity, let us consider the one with respect to $X^{*}_{0,+}$, which enters into the equation of motion of $X_{0,+}$. In this case, an explicit computation gives
\begin{equation}
\begin{split}
     \frac{\delta}{\delta X^{*}_{0,+}}\mathcal{U}[\vec{\psi}]&= \lim_{\epsilon \to 0}\frac{1}{\epsilon}\int dx [(X^*_{0,+}+\epsilon \delta(x-x^{'} )) X^*_{0,-}X_{1,+}X_{1,-} - X^*_{0,+} X^*_{0,-}X_{1,+}X_{1,-}]=\\
     &=\lim_{\epsilon \to 0}\frac{1}{\epsilon} \int dx \epsilon\, \delta(x-x^{'})\, X^*_{0,-}X_{1,+}X_{1,-}= X^*_{0,-}(x^{'},t)X_{1,+}(x^{'},t)X_{1,-}(x^{'},t).
\end{split}
\end{equation}
The contributions entering into the equations of motion for the other excitonic components are computed in the very same way. In particular, for what concerns all the numerical simulations shown in the main text, we set the parameter value $g_{xx}=1$ meV$\cdot\mu$m in our effective 1D model, without loss of generality (see the following discussion).
\subsection{Additional numerical results}\label{sm_sec:additional_numerical_results}
\begin{figure}[t]
    \centering
    \includegraphics[width=\textwidth]{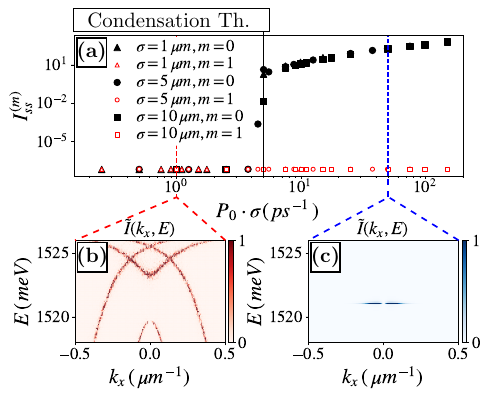}
    \caption{\textbf{(a)} Integrated steady state emission $I^{(m)}_{ss}$ associated to the modes $m=0$ and $m=1$ for different values of the spot size (see the legend). \textbf{(b)} Example of steady state emission spectrum below condensation threshold. \textbf{(c)} Example of steady state emission spectrum above condensation threshold ($\sigma=10\,\mu m)$. }
    \label{sm_fig:emission_U0_g0}
\end{figure}

In this section we show numerical results complementary to those shown in the main text. First, we report the behavior observed for both the integrated emission and the emission spectra in the case where both the linear coupling $U_{0,1}$ and the exciton-exciton nonlinearity $g_{xx}$ are set to zero in our simulations. In particular, the results are reported in Fig.~\ref{sm_fig:emission_U0_g0} for different values of the spot size, i.e.,  $\sigma = 1,5,10$ $\mu$m, respectively. Evidently, no second threshold in the emission from the $m=1$ component is seen in the numerical solutions, which complements the results reported in the main text: the presence of a second threshold is attributed to the exciton-exciton nonlinearity accounted by the interaction term $\mathcal{U}$ defined in Eq.~\ref{sm_eq:definition_nonlinearity}.

\begin{figure}
    \centering
    \includegraphics[width=\textwidth]{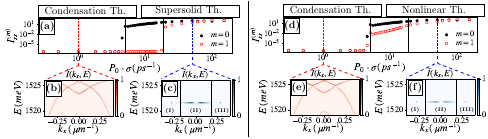}
    \caption{Numerical results obtained {for $g_{xx}\neq 0$ and $U_{0,1}=0$} (left panels), showing \textbf{(a)} the steady-state photonic emission, $I^{(m)}_{ss}$ ($m=0,1$), as a function of the injection rate, $P_0\cdot \sigma$ ($\sigma=5\,\mu m$), \textbf{(b)} the spectrally resolved emission normalized to the peak intensity, i.e. $\tilde{I}(k_x,E)=I(k_x,E)/max(I(k_x,E))$, for $P_0\cdot \sigma=1$ ps$^{-1}$ (i.e., below condensation threshold), and \textbf{(c)} for $P_0\cdot \sigma=50$ ps$^{-1}$ (above the supersolid transition threshold). In panels \textbf{(d)},\textbf{(e)},\textbf{(f)} we report the same quantities as in \textbf{(a),(b),(c)}, respectively, but assuming {a linear coupling matrix element $\hbar U_{0,1}=1$ meV}. We notice that in panels (c) and (f), the spectrally resolved emission in normalized to each peak in zones I (III) and II, respectively. The actual ratio between peak emission in region I (III) and II is about 4.}
    \label{sm_fig:emission_spot5}
\end{figure}

For what concerns the onset of polariton condensation (see Fig.~\ref{sm_fig:emission_U0_g0}a), numerical results describing $I^{(0)}_{ss}$ and obtained for different spot size nicely display the very same trend, when represented as a function of the injection rate, $P_0\cdot \sigma$, with a threshold value compatible with $5 ps^{-1}$ for our choice of the parameters. In addition, and in agreement with the discussion provided in the main text, even though the model assumes the presence of a coupling of $m=1$ modes with the active reservoir, the integrated steady state emission from this components, $I^{(1)}_{ss}$, is constant and compatible with the contribution associated only to the presence of noise $\vec{\xi}(x,t)$. In particular, the overall emission spectra below (i.e., Fig.~\ref{sm_fig:emission_U0_g0}b) and above  (i.e., Fig.~\ref{sm_fig:emission_U0_g0}c) condensation threshold, display the expected polariton band structure as well as the macroscopic occupation of the first gap-confined mode, respectively. However, no side peaks associated to emission from the $m=1$ branches at finite momentum that are isoenergetic with the condensate can be observed above threshold, as it was indeed the case in the results shown in Figs. 2c anb 2d in the main text.\\
{We now discuss the numerical results displaying the integrated emission as well as sample emission spectra at either low and very high injection rates for a specific value of the spot size $\sigma$. In the manuscript, we have shown similar results obtained for $\sigma=10$ $\mu$m, while here we report the outcome from an input source with $\sigma=5$ $\mu$m, for completeness. Similarly to what shown in Fig.~2 of the main text, we compare in Fig.~\ref{sm_fig:emission_spot5} the system response for $\hbar U_{0,1}=0$ and $\hbar U_{0,1}=1$ 1 meV, respectively. We notice that Figs.~\ref{sm_fig:emission_spot5}a and \ref{sm_fig:emission_spot5}d show a very similar trend of $I^{(0)}_{ss}$ versus $P_0\cdot \sigma$, as compared to Figs. 2a and 2b of the manuscript. In particular, the onset of condensation into the $m=0$ mode occurs at a very similar $P_0 \cdot \sigma$ value, both for $\sigma=5$ $\mu$m and $\sigma=10$ $\mu$m. On the other hand, a direct comparison between integrated emission from the $m=1$ mode, $I^{(1)}_{ss}$, shows that the real space spot size produces a shift of the supersolid/nonlinear thresholds (red dots in both Figs.~\ref{sm_fig:emission_spot5}a and \ref{sm_fig:emission_spot5}d), and a different coupling between $m=0$ and $m=1$ polaritons. Both issues can be understood by looking at the spectrally resolved emission patterns reported in Figs.~\ref{sm_fig:emission_spot5}c and \ref{sm_fig:emission_spot5}f. In particular, by reducing $\sigma$ to 5 $\mu$m, the momentum space broadening of the condensate (region II in Figs.~\ref{sm_fig:emission_spot5}c and \ref{sm_fig:emission_spot5}f) increases, i.e, the resulting condensate in the $m=0$ mode has longer tails as a function of $k_x$. This affects both the effective polariton-polariton interaction (see discussion in Sec.~\ref{sm_sec:instability}) and the coherent coupling between $m=0$ and $m=1$ modes due to the $U_{0,1}$ matrix element.}

\subsection{Scaling analysis of Mean-field solutions}\label{sm_sec:scaling}
\begin{figure}[t]
    \centering
    \subfigure[Below supersolid threshold]{
    \includegraphics[width=0.9\textwidth]{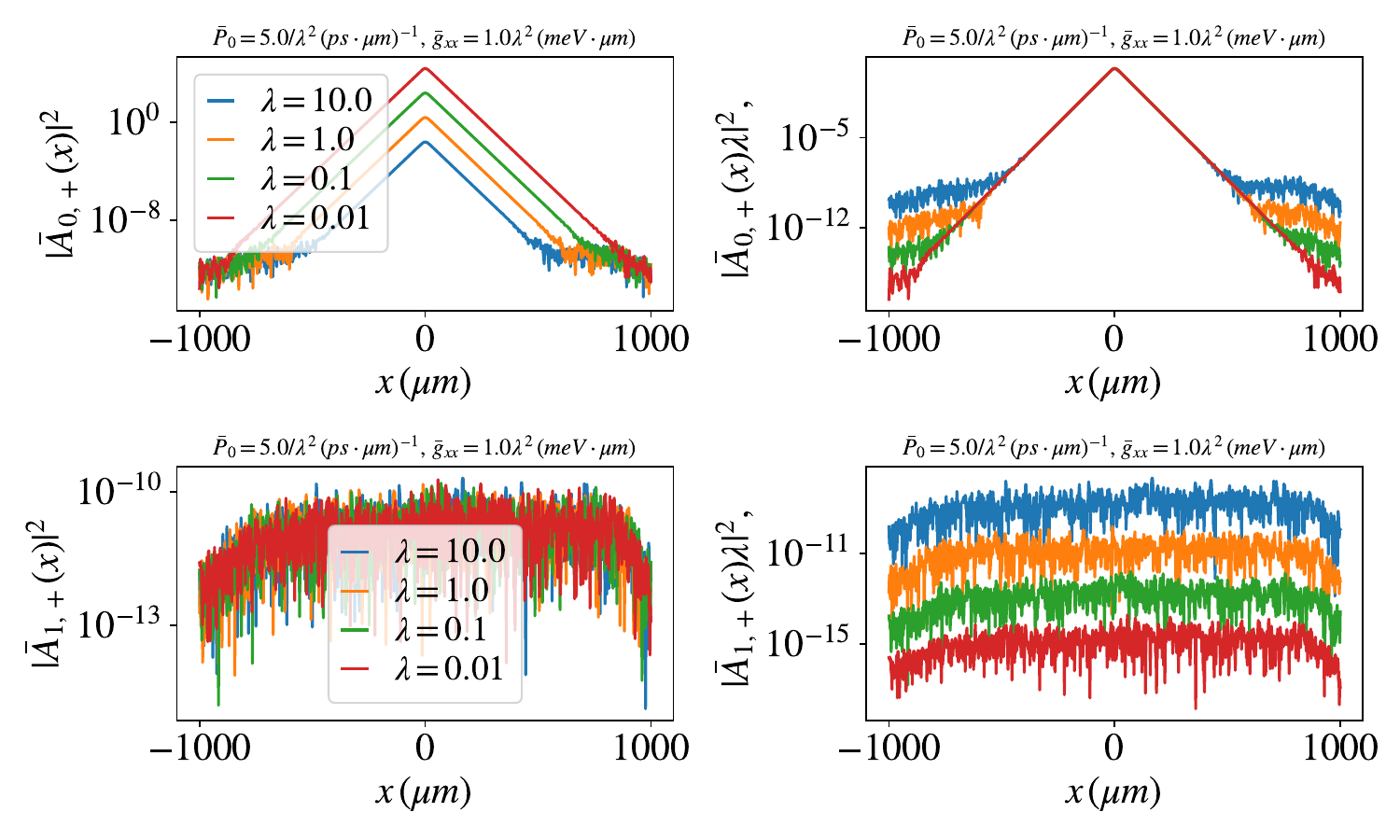}\label{sm_subfig:below}}
    \subfigure[Above supersolid threshold]{
    \includegraphics[width=0.9\textwidth]{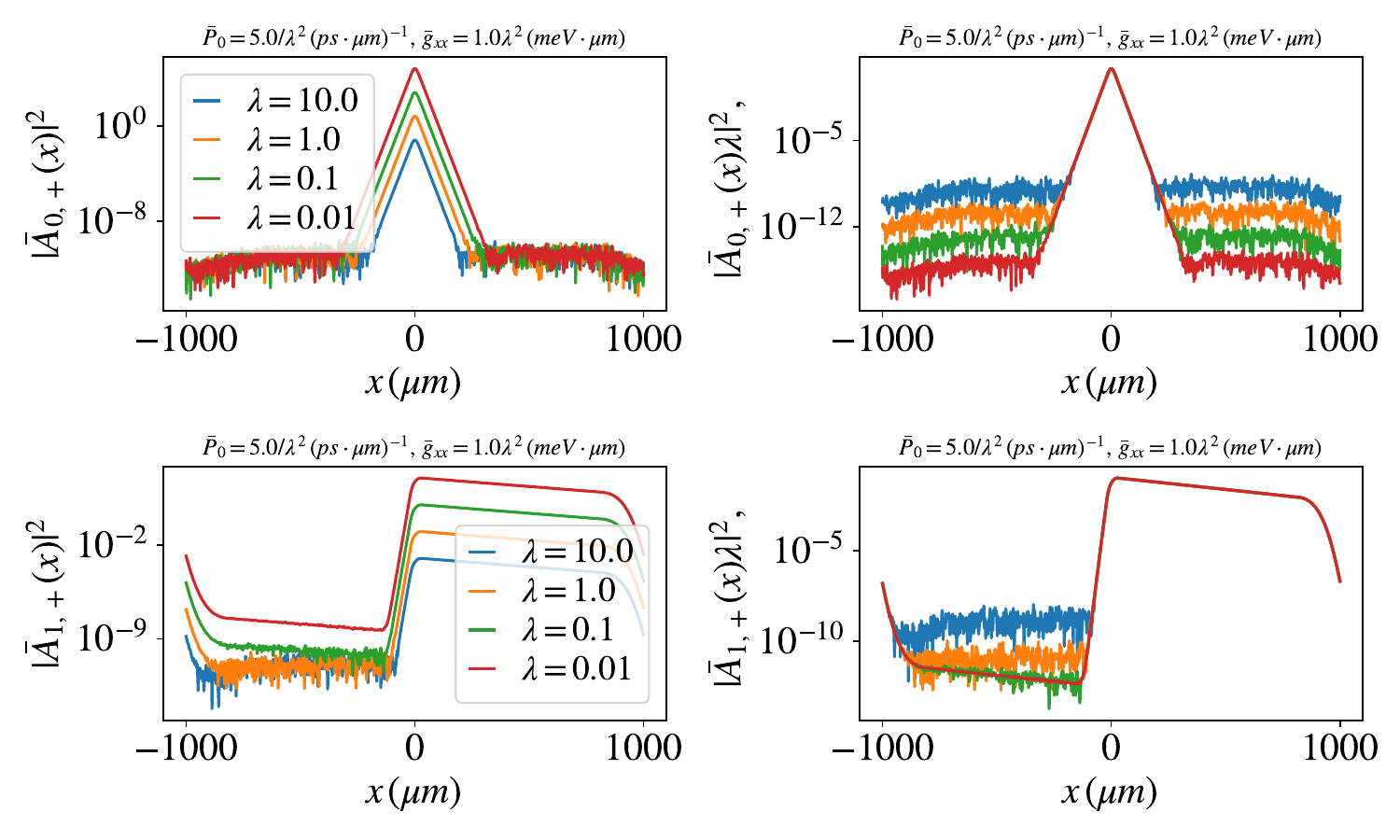}\label{sm_subfig:above}}
    \caption{Examples of scaling behavior above condensation threshold: {\bf (a) Behavior below the supersolid threshold}, reporting in the left panels the numerical results obtained for different values of the scaling factor ($\lambda$) when applying the parameters change as in Eq.~\ref{sm_eq:parameter_scaling}; in the right panels, we report the corresponding rescaled density. We stress that the $(1,+)$ component is only populated by noise, in this case. {\bf (b) Behavior above supersolid threshold}, reporting again in the left panels the numerical results obtained when scaling the parameters as prescribed in Eq~\ref{sm_eq:parameter_scaling}; the corresponding rescaled density is reported in the right panels. We notice that in this latter case a significant population is transferred to the $(1,+)$ mode. We notice that since $l=+$ modes are propagating configurations with positive group-velocity they travel across the sample from the left ($x<0$) to the right ($x>0$). This is the reason for the exponential decay observed at $x\ll 0$ (see, e.g., the last section of this file for further details on the model parameters employed).
    }
    \label{sm_fig:scaling_behavior}
\end{figure}

Here we briefly discuss the behavior of the set of coupled differential Eqs.~\ref{sm_eq:inactive_reservoir_eom}, \ref{sm_eq:active_reservoir_eom}, and \ref{sm_eq:psi_eom} under parameters scale transformation, which is relevant to quantitatively assess and interpret the polariton condensation threshold, also in connection with experiments. Let us suppose to have a solution of the dynamical system given by the equations of motion above, which essentially describes the behavior of $n_I$, $n_A$, and $\vec{\psi}$ for a given pump profile given by $P(x)=P_0\exp\{-(x^2)/(2\sigma^2)\}$. In addition, this solution arises from the  polariton band structure, as well as from the whole set of coupling parameters, i.e., $g$, $g_{xx}$, $w_I$, and $w_A$, respectively. Let us now suppose that we are interested in finding a solution corresponding to a different value of the exciton-exciton interaction constant, i.e., $\bar{g}_{xx}$. Without any loss of generality, we can think of $\bar{g}_{xx}$ as the result of a scaling transformation, which we define as
\begin{equation}
    \bar{g}_{xx}=\lambda^2 g_{xx} \, ,
\end{equation} 
for some scalar and dimensionless value, $\lambda$. The action of such a scaling can be essentially reabsorbed into the set of coupled equations by performing the following change of variables:
\begin{equation}\label{sm_eq:parameter_scaling}
    \left(P_0,\, n_{I},\,n_{A},\,\vec{\psi},\, g,\,w_I,\,w_A\right) \longmapsto\left(\bar{P}_0,\, \bar{n}_{I},\,\bar{n}_{A},\,\vec{\bar{\psi}},\, \bar{g},\,\bar{w}_I,\,\bar{w}_A\right)=  \left(\frac{P_0}{\lambda^2},\, \frac{n_{I}}{\lambda^2},\,\frac{n_{A}}{\lambda^2},\,\frac{\vec{\psi}}{\lambda},\, g \lambda^2,\,w_I \lambda^2,\,w_A \lambda^2\right) \, .
\end{equation}
In particular, this parameters rescaling is quite useful when considering the behavior above condensation threshold. Numerical results displaying the scaling of the density associated to the steady state $\bar{A}_{m,l}(x)$ (for the $0,+$ and $1,+$ modes, respectively) are reported in Fig.~\ref{sm_fig:scaling_behavior}. A selection of numerical results showing the system response below the supersolid threshold is explicitly reported in Fig.~S\ref{sm_subfig:below}. Here, it is possible to notice (see, e.g., the top row), by direct comparison of left and right panels, that once after proper rescaling of the numerical results obtained for different values of $\lambda$, the outcome is essentially coincident in the same condensate profile, within numerical accuracy. In particular, in this case the $(m,l)=(1,+)$ mode (as well as all the other components associated to mode $m=1$) are just populated by noise. A similarly good agreement is observed above the supersolid threshold, as shown in Fig.~S\ref{sm_subfig:above}. 

\subsection{Instability analysis}\label{sm_sec:instability}
\begin{figure}
    \centering
    \includegraphics[width=0.8\textwidth]{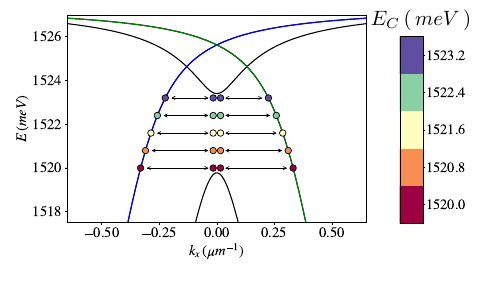}
    \caption{Sketch of the resonant processes relevant for the emergence of a polariton supersolid phase. Different colors are used to identify processes corresponding to a condensate appearing in the gap at different energy, here identified as $E_C$. Above the supersolid transition threshold, two particles belonging to the $m=0$ polariton condensate around zero-wavevector (central dots) are efficiently scattered into a pair of particles with opposite $k_x$ and same energy $E_C$ (lateral dots) on the side $m=1$ polariton branches.}
    \label{sm_fig:example_processes}
\end{figure}

We hereby perform a linear stability analysis, concerning the dynamical equations around the condensate configuration, with the aim of theoretically assessing  the conditions for which the polariton condensate becomes unstable in favour of the supersolid phase.\\
Owing to the form of the exciton-exciton interaction term, and since no significant population is stored in the $m=1$ subspace below the supersolid phase transition threshold, the non-trivial part of the dynamics is encoded into the following reduced set of coupled equations:
\begin{equation}
    \frac{d}{dt}\left(
    \begin{array}{c}
         A_{1,+}\\
         X_{1,+}\\
         A_{1,-}\\
         X_{1,-}
    \end{array}
\right)=
-\frac{i}{\hbar} \left(
\begin{array}{cc|cc}
     E_+(x)& \hbar \Omega_R & 0 & 0 \\
     \hbar \Omega_R  & E_X & 0 & 0 \\
     \hline
     0&0 &E_-(x)& \hbar \Omega_R \\
     0&0&\hbar \Omega_R  & E_X  
\end{array}
\right)\left(
    \begin{array}{c}
         A_{1,+}\\
         X_{1,+}\\
         A_{1,-}\\
         X_{1,-}
    \end{array}
\right)
-\left(
    \begin{array}{c}
         \gamma_{ph} A_{1,+}\\
         \gamma_X X_{1,+}\\
         \gamma_{ph} A_{1,-}\\
         \gamma_X X_{1,-}
    \end{array}
\right)
-\frac{i}{\hbar}g_{xx} f(x) \left(
    \begin{array}{c}
         0\\
         X^*_{1,-}\\
         0\\
         X^*_{1,+}
    \end{array}
\right)e^{-i2\frac{E_C t}{\hbar}}
\end{equation}
where $f(x)\equiv X^{(ss)}_{0,+}(x)X^{(ss)}_{0,-}(x)$ and $E_C$ account for the excitonic components of the polariton condensate and its energy (i.e., corresponding to the dispersionless gap-confined mode), respectively. Due to the time dependence introduced by the condensate, only  specific points on the resulting polariton dispersion can be resonantly coupled. This can be seen by taking the following steps:
\begin{itemize}
    \item move into wave vector space by defining the Fourier transforms
\begin{equation}
    \left(
    \begin{array}{c}
         A_{1,+}(x,t)\\
         X_{1,+}(x,t)\\
         A_{1,-}(x,t)\\
         X_{1,-}(x,t)
    \end{array}
\right)=\int \frac{dk_x}{2\pi}e^{ik_x x}\left(
    \begin{array}{c}
         A_{1,+}(k,t)\\
         X_{1,+}(k,t)\\
         A_{1,-}(k,t)\\
         X_{1,-}(k,t)
    \end{array}
\right);
\end{equation}
\item diagonalize the free part of the Hamiltonian in order to determine the unitary matrix to change the basis (i.e., the $2\times 2$ $U_k$ matrix) and project the evolution of photons and excitons onto polaritons, both into the propagating and counterpropagating sectors. By defining  $ P_{U,l}(k,t)$ and $ P_{L,l}(k,t)$, respectively the upper and lower polariton branches, we have 
\begin{equation}
\left(
    \begin{array}{c}
         A_{1,\pm}(k,t)\\
         X_{1,\pm}(k,t)\\
    \end{array}
\right)=U^{(\pm)}_k\left(
    \begin{array}{c}
         P_{U,\pm}(k,t)\\
         P_{L,\pm}(k,t)\\
    \end{array}
\right),\quad U^{(\pm)}_k=
\left(
\begin{array}{cc}
    a^{(\pm)}_{U}(k) & a^{(\pm)}_{L}(k) \\
    x^{(\pm)}_{U}(k) & x^{(\pm)}_{L}(k)
\end{array}
\right).
\end{equation}
In particular, $a^{(\pm)2}_{p}(k)$ and $x^{(\pm)2}_{p}(k)$ ($p=U,L$, defining upper and lower polariton branches, respectively) denote the photonic and excitonic fractions of the (upper and lower) polariton amplitudes, respectively;

\item consider only contributions simultaneously preserving energy and the momentum, which in this case correspond to the processes connecting the condensate (close to $k_x=0$) to the $m=1$ modes at the same energy (as schematically shown, e.g., in Fig.~\ref{sm_fig:example_processes}).
\end{itemize}

After some tedious algebra, by neglecting non-resonant terms, we finally obtain the following set of coupled differential equations
\begin{equation}\label{sm_eq:approximate_dynamics}
    \frac{d}{dt}\left(
\begin{array}{c}
     \tilde{P}^{(+)}_{L} (k) \\
     \tilde{P}^{(-)*}_{L} (-k) 
\end{array}
    \right)\simeq \left(
    \begin{array}{cc}
        -\Gamma^{(+)}_{L,L}(k) & -i g^{(eff)}_{xx}x^{(+)}_{L}(k)x^{(-)}_{L}(-k) \\
        i g^{(eff)}_{xx}x^{(-)}_{L}(-k)x^{(+)}_{L}(k) &-\Gamma^{(-)}_{L,L}(-k)   \\
    \end{array}
    \right) \left(
\begin{array}{c}
     \tilde{P}^{(+)}_{L} (k) \\
     \tilde{P}^{(-)*}_{L} (-k) 
\end{array}
\right) \, ,
\end{equation}
in which $\Gamma^{(\pm)}_{L,L}(k)=\gamma_{ph}a^{(\pm)2}_{L}(k)+\gamma_{X} x^{(\pm)2}_{L}(k)$ and with $g^{(eff)}_{xx}=\frac{g_{xx}}{2\pi \hbar}\hat{f}(0)$ is the effective exciton-exciton interaction that is controlled by the zero-wavevector component of the $f(x)$ Fourier transform. 
The terms $\tilde{P}^{(\pm)}_{L} (k)$ represent the rotated amplitudes of lower polariton branches, i.e., 
\begin{equation}
    \tilde{P}^{(\pm)}_{L} (k)= e^{i E^{(\pm)}_{L}(k) t/\hbar}P^{(\pm)}_{L} (k) \quad (\mbox{with}\, E^{(\pm)}_{L}(k)=E_C)
\end{equation}
with 
 \begin{equation}
    E^{(\pm)}_{L}(k)=\frac{E_X+ E_{1,\pm}(k)}{2}-\frac{1}{2}\sqrt{(E_X- E_{1,\pm}(k))^2+4 (\hbar \Omega_{R,1})^2} \, ,
\end{equation}
which describe the energy-momentum dispersion of the lower polariton branches.

Since the evolution of $\tilde{P}^{(\pm)}_{L} (k)$ is linearly coupled to the evolution of $\tilde{P}^{(\mp)*}_{L} (-k)$, these two amplitudes are defined up to an identical phase. As a consequence, $\tilde{P}^{(\pm)}_{L} (k)$ and $\tilde{P}^{(\mp)}_{L} (-k)$ are defined up to a pair of conjugated phases (whose values become pinned once the system experiences a transition to the supersolid phase, due to the presence of a random fluctuation introduced by the noise). In other words, for any value of the phase $\phi_+$, the equation of motions reported in Eq.~\ref{sm_eq:approximate_dynamics} are invariant under the following phase transformation
\begin{equation}
    \tilde{P}^{(+)}_{L} (k)\to e^{i\phi_+}\tilde{P}^{(+)}_{L} (k),\quad \tilde{P}^{(-)}_{L} (-k)\to e^{i\phi_-}\tilde{P}^{(-)}_{L} (-k),
\end{equation}
provided that $\phi_+ + \phi_-=0$. In particular, the phases $\phi_+$ and $\phi_-$ correspond to $\phi_L$ and $\phi_R$ of the main text, respectively. Here, we avoided using $L/R$ as a subscript for the phases to make clear distinction with the subscript $L$ used for lower polariton states.

\begin{figure}
    \centering
    \includegraphics[width=0.5\textwidth]{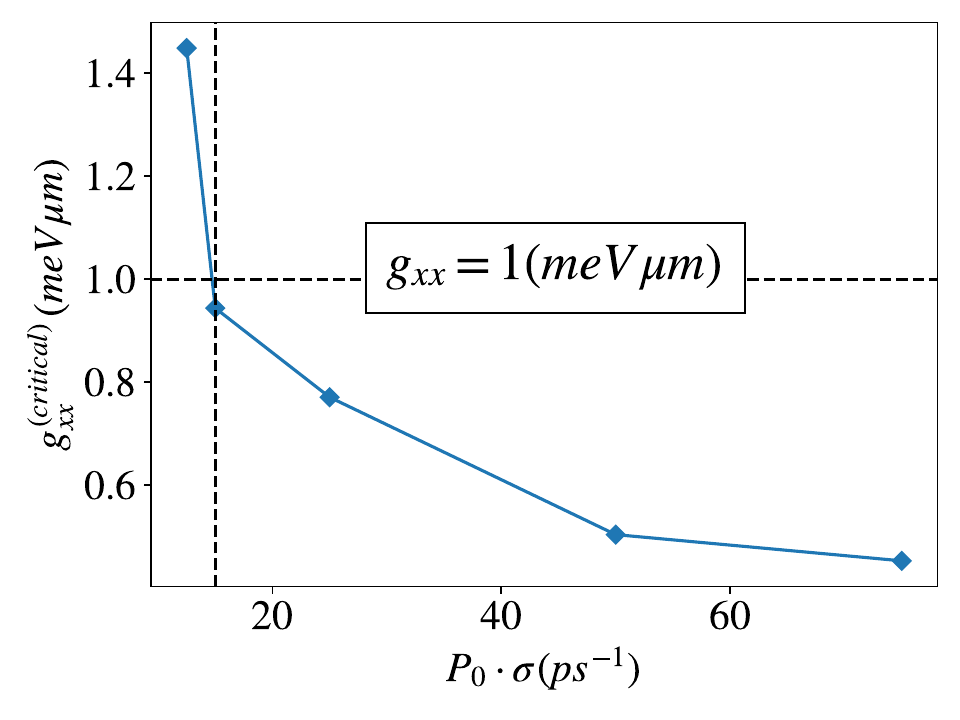}
    \caption{Behavior of $g^{(critical)}_{xx}$ (see Eq.~\ref{sm_eq:instability_condition}) as function of the injection rate, $P_0\cdot \sigma$. The horizontal dashed line corresponds to the value of $g_{xx}$ used in this work. The vertical line located in the neighborhood of $P_0\cdot \sigma\simeq 15\, ps^{-1}$ provides an estimate for the onset of supersolid transition.}
    \label{sm_fig:instability_plot}
\end{figure}

As a final result, we obtain that the condensate becomes linearly unstable in favour of the supersolid configuration when the set of coupled equations in Eq.~\ref{sm_eq:approximate_dynamics} is characterized by an exponentially growing solution in time. By looking at the eigenvalues of the coefficient matrix, this is analytically found to occur when
\begin{equation}\label{sm_eq:instability_condition}
    g_{xx}\geq g^{(critical)}_{xx}=\frac{2\pi \hbar}{\vert \hat{f}(0)\vert}\sqrt{\frac{\left(\Gamma^{(+)}_{L,L}(k)\right)^2+\left(\Gamma^{(-)}_{L,L}(-k)\right)^2}{2 x^{(+)2}_{L}(k) x^{(-)2}_{L}(-k)}} \, .
\end{equation}
A comparison between the expression reported in Eq.~\ref{sm_eq:instability_condition}, the data reported in the main text in Fig. 2a, and the particular value of the exciton-exciton nonlinearity $g_{xx}$ used in the present analysis is reported in Fig.~\ref{sm_fig:instability_plot}. The results obtained by means of the linear stability analysis provide an estimate for the supersolid transition threshold ($P_0\cdot \sigma\simeq 15\, ps^{-1}$, vertical dashed line), which is in good agreement with the numerical results reported in the main text.\\

\begin{figure}
    \centering
    \includegraphics[width=0.9\textwidth]{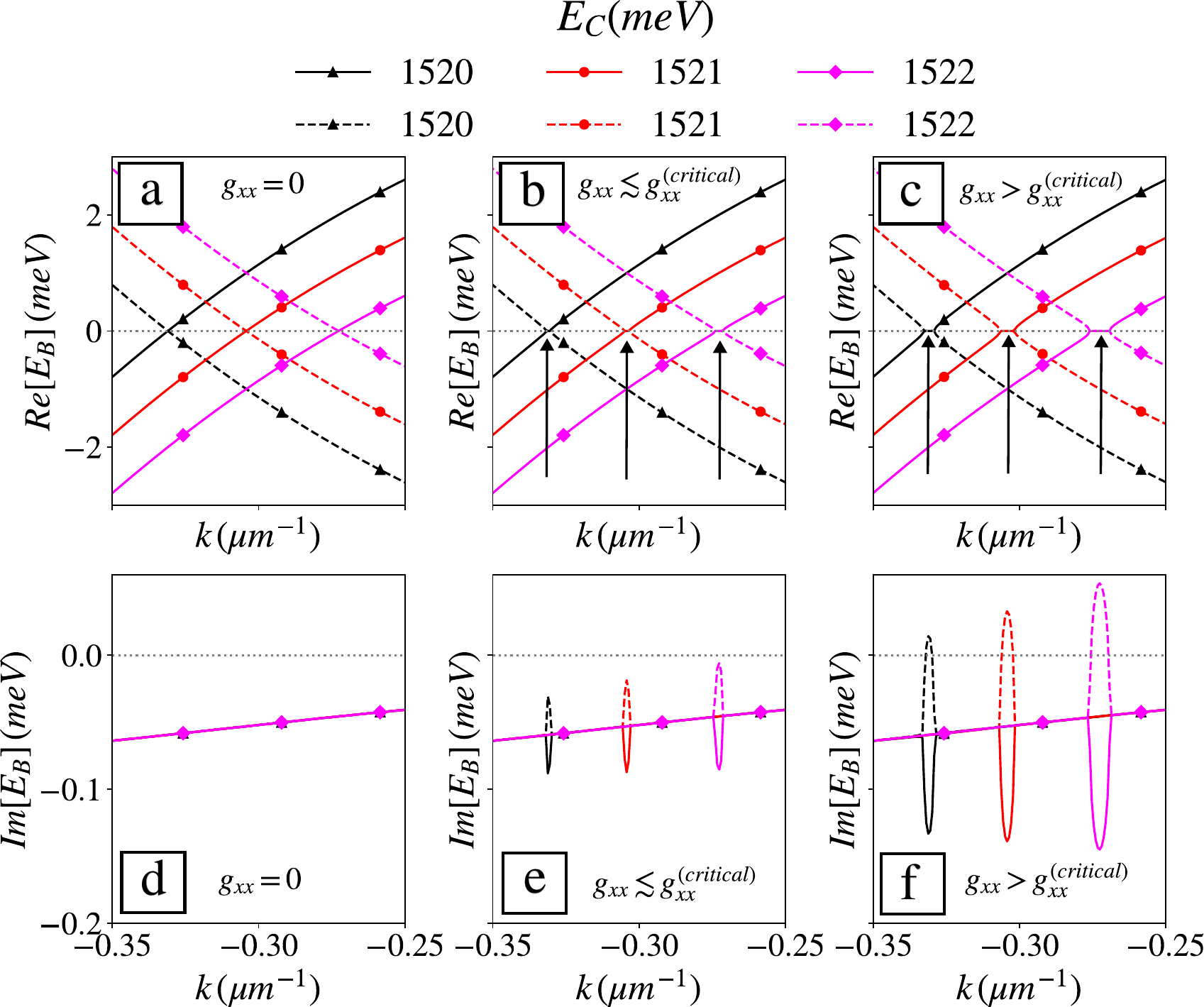}
    \caption{\DN{Panels (a) and (d): real and imaginary parts of the Bogoliubov spectrum $E_B(k)$ for zero nonlinearity. Panels (b) and (e): real and imaginary parts of the Bogoliubov spectrum $E_B(k)$ below the onset of parametric instability. Panels (c) and (f): real and imaginary parts of the Bogoliubov spectrum $E_B(k)$ above the onset of parametric instability. The different curves have been obtained for the three different values of the energy of the condensate specified in the top legend. }}
    \label{sm_fig:bogo_bands}
\end{figure}

\DN{Another relevant issue that can be addressed by means of the linear stability analysis is whether or not excitation created on top of the polariton condensate preserve the superfluid nature expected for a proper supersolid. According to the analysis provided in \cite{Grudinina2023}, the polariton condensate formed in the type of structures considered in this work are characterized by a finite speed of sound along the grating direction. This means that impurities moving slower than sound, in general, cannot perturb the superfluid character of the system. The same idea applies to polaritons generated due to the scattering from modes $m=0$ to modes $m=1$. At first one might think that, since $m=1$ polariton modes resonating with the condensate at $E_C$ are characterized by a group-velocity larger than the typical speed of sound determined in Ref \cite{Grudinina2023}, the scattering of particles from the condensate to such branches would in general destroy any superfluid character present in the condensate. We argue that this is not the case. Indeed, due to the parametric nature of the instability leading to the formation of the patterns observed above the supersolid threshold, the modulation created on top of the condensate is given by the superposition of a propagating state ($\tilde{P}^{(\pm)}_{L} (k)$) and a counterpropagating one ($\tilde{P}^{(\mp) *}_{L} (k)$): excitations formed in such a way are characterized by a flat energy-wavector band, corresponding to zero group-velocity regions \cite{Carusotto_Ciuti_RMP2013}, which precurse a stationary modulation at rest. Such a fact can be directly verified by looking at the Bogoliubov spectrum that can be extracted from the linear stability analysis around the un-modulated stationary state. An example of such spectra for zero exciton-exciton nonlinearity ($g_{xx}=0$, panels (a,d)), for a nonlinearity below the critical value for supersolidity ($g_{xx}\lesssim g^{(critical)}_{xx}$, panels (b,e)) and above it ($g_{xx}>g^{(critical)}_{xx}$, panels (c,f)) is shown in Fig. \ref{sm_fig:bogo_bands}. For each value of the condensate energy $E_c$ (see the top legend), we determined the Bogoliubov spectrum by looking at the eigenvalues of the coefficient matrix appearing in the right-hand side of Eq. \ref{sm_eq:approximate_dynamics} as function of $k$. In particular, in panels (a), (b) and (c) the zero of the energy axis is assumed to correspond to the condensate energy $E_c$. With such a choice, the $E_{B}(k)$ dispersion accounts for the energy of excitations with respect to the condensate energy.
In the linear case, \emph{i.e.} zero nonlinearity (panels a and d), the Bogoliubov spectrum is equivalent to that of the propagating branch and the “ghost” counterpropagating one, associated respectively to $\tilde{P}^{(\pm)}_{L} (k)$ and $\tilde{P}^{(\mp) *}_{L}(k)$. In particular, in this case the solid-line and dashed-line curves correspond to $Re[E_B(k)]= E_{L}^{(+)}(k)-E_c$ and $Re[E_B(k)]= -E_{L}^{(-)}(-k)+E_c$, respectively.
As the nonlinearity is introduced in the system evolution, the two branches mentioned here above hybridize, leading to the formation of a plateau with zero energy, \emph{i.e.} $Re[E_B(k)]=0$, and zero group-velocity in the energy-wavevector space. Such plateaux are visible both in panels (b) and (c), in the regions indicated by the vertical arrows. Furthermore, in correspondence of such flat band regions also the loss content of the Bogoliubov dispersion gets modified. Such a fact is observed in panels (e) and (f), where clear peaks/dips are visible in the two Bogoliubov branches. In particular, as expected, when the nonlinearity exceeds the critical value expected for the parametric process, one of the two Bogoliubov dispersions is characterized by $Im[E_B(k)]>0$, typical signature of the onset of an instability.}\\
To conclude this Section, it is worth pointing out a few comments related to the expression reported in Eq.~\ref{sm_eq:instability_condition}:
\begin{itemize}
    \item[i)] the critical exciton-exciton interaction leading to the supersolid phase transition, $ g^{(critical)}_{xx}$, is proportional to $(\vert \hat{f}(0)\vert)^{-1}$, i.e., the larger the condensate occupation, the smaller the exciton interaction that is required to trigger the instability;
    \item[ii)] $ g^{(critical)}_{xx}$ is also proportional to the loss rates of the $m=1$ polaritons at finite wave vector: the larger the linewidth of these polariton branches, the more unlikely is to trigger a stable scattering towards these modes (i.e., particle loss dominates over building up of a supersolid phase);
    \item[iii)] $ g^{(critical)}_{xx}$ is further inversely proportional to the excitonic fractions of the $m=1$ polariton eigenmode: since the resonant nonlinear scattering from the condensate to the finite momentum modes is ultimately triggered by exciton-exciton interactions, the smaller the excitonic fraction in these states, the larger the bare exciton interaction that is necessary to sustain the stable scattering from the condensate to $m=1$ states. 
\end{itemize}

\subsection{Numerical details}\label{sm_sec:numerical_details}
\subsubsection{Spatial grid and absorbing potential}
Spatial dependence of photonic and excitonic amplitudes has been represented by considering a 1D box extending from $-L$ to $L$, with  $L=1000$ $\mu$m (as compatible with realistic samples). The number of points used in the spatial discretization mesh is given by $2^{10}+1$. The large size of the box is needed to acquire resolution in momentum space around $k_x=0$, to capture the emission features associated to the dark photonic mode. In order to suppress finite size effects from the box boundaries (i.e., particle reflection from the left/right boundaries towards the center of the box), we included an imaginary parabolic potential extending for $\vert x \vert\geq 800\, \mu m$. The extra damping contributions introduced by this potential are visible, e.g., in Fig.~\ref{sm_fig:scaling_behavior}, where population decay within the $\vert x \vert\geq 800$ $\mu$m region is clearly evidenced.

\subsubsection{Noise term and integration scheme}
Time evolution has been performed by perturbing the deterministic evolution of the reservoirs-polariton dynamical system with a Gaussian-correlated noise term in the equations of motion (see Eq.~\ref{sm_eq:psi_eom}), $\vec{\xi}(x,t)$. In this work, we employed a standard four-point explicit Runge-Kutta (RK4) scheme with a time step given by $dt=10^{-3}$ (absolute time scale $\tau=10\,ps$, corresponding to an energy scale: $\Delta E=\hbar/\tau \sim 0.066$ meV). In particular, at each step of the time evolution, $t_n$, the following algorithms allows to evolve the coupled equations in time:
\begin{itemize}
    \item [1)] Runge-Kutta step: $(n_I(x,t_n),n_A(x,t_n),\vec{\psi}(x,t_n))\to (n_I(x,t_n+dt),n_A(x,t_n+dt),\vec{\psi}(x,t_n+dt))$;
    \item [2)] Generation of the noise: for each point $x_j$ of the grid, we generated $\vec{\xi}(x_j,t_n)$ as $\vec{\xi}(x_j,t_n)=\vec{\xi}_R+ i \vec{\xi}_I$, where the two arrays of (pseudo)random numbers, $\vec{\xi}_R$ and $\vec{\xi}_I$, were obtained by sampling a Gaussian distribution with zero mean and standard-deviation $\sqrt{dt/2}$;
    \item [3)] Perturbation of the polariton subsystem: we added to $\vec{\psi}(x_j,t_n+dt)$ the noise perturbation $\delta \vec{\psi}\equiv \eta \,\vec{\xi}(x_j,t_n)$, with a scaling parameter $\eta=10^{-4}$.
\end{itemize}
We notice that the presence of noise is necessary to obtain meaningful numerical evolution, in particular, it allows to make the trivial initial configuration with zero excitations in the system ($\vec{\psi}(x,t)=\vec{0}$) unstable.
%(whenever the dynamical system is characterized by instabilities). 
Hence, the numerical time evolution has been performed by iterating such a procedure for a total number of time steps, $N_t$, given by $N_t=100\cdot 2^{12}$. In particular, the 300 different steady-state configurations employed in the main text to perform the statistical analysis have been obtained by evolving an initially empty configuration in the presence of 300 different randomly generated sequences of noise arrays.  
\bibliography{bibliography}

\end{document}